\newif\ifconfver
\confverfalse      %declaring conference version false
\newif\ifcutshort      %this level shortens the equations
\cutshorttrue

\newif\ifcutshortlvltwo  %this level takes out some examples, figs., and sim.
\cutshortlvltwofalse

\ifconfver
\documentclass[10pt,twocolumn,twoside]{IEEEtran}
\else
\documentclass{IEEEtran}
\fi

\usepackage{ dsfont }
\usepackage{cite}
\usepackage{graphicx}
\usepackage{psfrag}
\usepackage{url}
\usepackage{stfloats}
\usepackage{amsfonts,amssymb,amsmath,bm}
\usepackage{array}
\usepackage{caption}
\usepackage{cases}
\usepackage{calc}

\usepackage{longtable}
\usepackage{stfloats}  % Written by Sigitas Tolusis
\usepackage{graphics,booktabs,color,epsfig,enumerate}
\usepackage{multirow}
\usepackage{algorithm,algpseudocode}
\usepackage{subfigure}
\usepackage{epstopdf}
\usepackage{array}
\usepackage{stfloats}
\usepackage{soul}
\newtheorem{lemma}{Lemma}
\newtheorem{theorem}{Theorem}
\newtheorem{fact}{Fact}
\newtheorem{assum}{Assumption}

{\begin{list}{}{
			\settowidth{\labelwidth}{\mbox{\textnormal{#1}}}%
			\setlength{\leftmargin}{\labelwidth+\labelsep}}}%
	{\end{list}}

\makeatletter
\def\changeBibColor#1{%
	\in@{#1}{}%  list of colored bib items
	\ifin@\color{blue}\else\normalcolor\fi
}

\def\blue {\color{blue}}

\begin{document}
	
	%	\allowdisplaybreaks[4]
	
	\bibliographystyle{IEEEtran}

	\def\blue{\color{blue}}
	\def\red{\color{red}}
	\definecolor{orange}{RGB}{255,107,0}
	\def\orange{\color{orange}}

	% ==================================================
%	\title{Cross-Receiver  Radio Frequency Fingerprint Identification via Domain Adaptation}
	\title{Mitigating Receiver Impact on  Radio Frequency Fingerprint Identification via Domain Adaptation}
	
	\ifconfver \else {\linespread{1.1} \rm \fi
		
		\author{Liu Yang, Qiang Li, Xiaoyang Ren, Yi Fang and Shafei Wang
			\thanks{L. Yang, Q. Li and X. Ren are with
				School of Information and Communication  Engineering, University of Electronic Science and Technology of China, Chengdu, P.~R.~China, 611731. Q. Li is also affiliated with  Laboratory of Electromagnetic Space Cognition and Intelligent Control, Beijing, China, 100083. }
			\thanks{Y. Fang  and S. Wang are with Laboratory of Electromagnetic Space Cognition and Intelligent Control, Beijing, China, 100083. }
			%		\thanks{M. Shao is with Dept. Electronic Engineering, The Chinese University of Hong Kong, Hong Kong. E-mail: mjshao@ee.cuhk.edu.hk}
			\thanks{Q. Li is the corresponding author. E-mail: lq@uestc.edu.cn}
			\thanks{This work was
supported by the National Natural Science Foundation of China under Grant
62171110.}
		}
	
		\maketitle

		% Title.
		% ------
		%\title{Cross receiver radio frequency fingerprint identification}
		%
		% Single address.
		% ---------------
		%\name{Liu Yang$^{1}$,  Qiang Li$^{1,2}$, Xiaoyang Ren$^{1}$, Yi Fang$^{2}$}
		%\address{{$^{1}$School of Infor. \& Commun. Engineering, University of Electronic Science and Technology of China}\\
		%	$^{2}$Laboratory of Electromagnetic Space Cognition and Intelligent Control, Beijing 100089, China}
		%
		% For example:
		% ------------
		%\address{School\\
		%	Department\\
		%	Address}
		%
		% Two addresses (uncomment and modify for two-address case).
		% ----------------------------------------------------------
		%\twoauthors
		%  {A. Author-one, B. Author-two\sthanks{Thanks to XYZ agency for funding.}}
		%	{School A-B\\
		%	Department A-B\\
		%	Address A-B}
		%  {C. Author-three, D. Author-four\sthanks{The fourth author performed the work
		%	while at ...}}
		%	{School C-D\\
		%	Department C-D\\
		%	Address C-D}
		%
		%\begin{document}
		%	\topmargin=0mm
		%	\ninept
		%
		%	\maketitle
		%
		\begin{abstract}
			%		With the development of wireless communication, 
			
			Radio Frequency Fingerprint Identification (RFFI), which exploits   non-ideal hardware-induced unique distortion resident in the transmit signals to identify an emitter, is emerging as a means to enhance the security of communication systems. Recently, machine learning has achieved great success in developing state-of-the-art RFFI models. However, few works consider  cross-receiver RFFI problems, where the RFFI model is trained and  deployed on different receivers. Due to altered receiver characteristics, direct deployment of RFFI model on a new receiver leads to significant performance degradation. To address this issue, we  formulate the cross-receiver RFFI as a model adaptation problem, which adapts the trained model to unlabeled signals from a new receiver. We first develop a theoretical generalization error bound for the adaptation model. Motivated by the bound, we propose a novel method to solve the cross-receiver RFFI problem, which includes domain alignment and adaptive pseudo-labeling. The former aims at finding a feature space where both domains exhibit similar distributions, effectively reducing the domain discrepancy. Meanwhile, the latter employs a dynamic pseudo-labeling scheme to implicitly transfer the label information from the labeled receiver to the new receiver. Experimental results indicate that the proposed method can effectively mitigate the receiver impact and improve the cross-receiver RFFI performance.

			%		performance degradation caused by .

		\end{abstract}
		\section{Introduction}
		\label{sec:intro}

\begin{figure}[]
			\centering
			\centerline{\includegraphics[width=8.4cm]{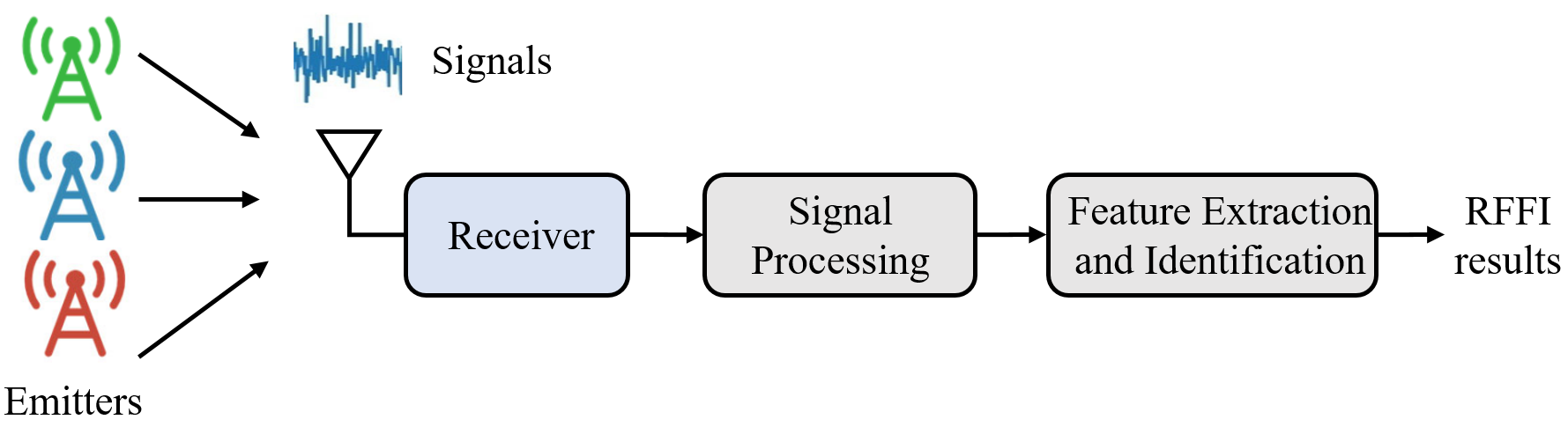}}
			%  \vspace{1.5cm}
			\caption{The block diagram of RFFI system.}
			\label{fig:RFFI_system}
			%		\vspace{-5pt}
		\end{figure}

		In the rapidly evolving wireless communication landscape, ensuring the security and trustworthiness of interconnected devices has become a critical concern. In this case, Radio Frequency Fingerprint Identification (RFFI) has emerged as a promising and innovative physical authentication technology. It introduces an additional method for device authentication at the physical layer, augmenting traditional authentication methods at higher layers, as evidenced by prior research~\cite{zhang2023radio, chen2019radio, tian2020identity}.
		RFFI is a technology that identifies RF emitters by analyzing the modulation introduced by non-ideal characteristics of the emitting device, also known as their unique radio frequency (RF) ``fingerprint''~\cite{xu2021adaptive}. 
		The RFFI system is shown in Fig.~\ref{fig:RFFI_system}, which mainly includes signal receiving, signal processing, and feature extraction. Existing researches mainly cover signal processing and feature extraction techniques applied to received signals in order to improve the recognition accuracy of emitters.
		Since the RFFI problem resembles a classification task with the received RF waveform as input and the corresponding emitter class as output, it is popular to employ machine learning approaches,  especially deep learning (DL), to train an end-to-end classification model to extract RF fingerprints features and classify them through supervised learning. Recent studies reveal that the DL-based RFFI approaches have outperformed the traditional signal processing ones and attained state-of-the-art recognition performance~\cite{zhang2023radio, liu2022radio, zhang2023data}. These data-driven techniques have the potential to revolutionize how to identify RF emitters by analyzing their unique fingerprints encoded within RF waveforms. 
		
		However, few studies have discussed the impact of receivers on RFFI systems in Fig.~\ref{fig:RFFI_system}. A critical challenge emerges as we seek to deploy the above DL-based RFFI models in real-world scenarios due to the impact of receivers.
		Imagine a scenario where we have meticulously crafted a training dataset with signals from a certain receiver, Rx-1, to train an end-to-end RFFI recognition model. This model accurately identifies emitters. However, Rx-1 could break down during its utilization, and be replaced by another receiver, Rx-2.
		The transition from Rx-1 to Rx-2 introduces a fundamental problem: the distinctive reception characteristics of these receivers lead to a data distribution shift. This shift, in turn, results in a dramatic drop in the recognition performance of the RFFI model trained exclusively with Rx-1 data when applied to Rx-2~\cite{merchant2019toward, shen2022towards}; see also the results in Section~\ref{sec:experiment}.
		
		One potential solution is apparent: we could re-sample emitters' waveforms and painstakingly label data collected by Rx-2, subsequently re-training the RFFI model with this new dataset. However, this approach not only demands substantial time and resources to re-create the training dataset from scratch, but also renders the original labeled data from Rx-1 obsolete. In recent years, transfer learning has garnered significant attention, particularly the pre-training and fine-tuning paradigm~\cite{devlin2018bert}. This approach allows models to be pre-trained on large-scale datasets and then fine-tuned on a related dataset with limited labels to achieve better performance. Despite its advantages, transfer learning still requires manual annotation of new data after it has been collected, which does not fully automate the adaptation to new data from a new receiver.
		In light of this, the following cross-receiver RFFI problem is put forward: 
		
		{\it Can we efficiently harness the original labeled data from Rx-1 and adapt the RFFI model to the distinct reception characteristics of Rx-2 without the need for manual labeling Rx-2's data? If yes, can we theoretically justify the performance gap between the adapted model and the optimal model that is trained directly with the labeled data from Rx-2?}
		
		%\subsection{Contribution of this work}
		%To overcome this challenge, we efficiently harness the original training data from Rx-1 and adapt the RFFI model to the distinct reception characteristics of Rx-2 without the need for manual labeling, which is called the cross-receiver RFFI problem. 

		While no formalization of this problem has been attempted, some early related research has been conducted.
		Merchant \emph{et al.}~\cite{merchant2019toward} advocated reducing the effect of receivers by minimizing the distance of the predicted probability from both receivers. Shen \emph{et al.}~\cite{shen2022towards} focused on learning receiver-independent features by adversarial training over multiple receivers’ data. Although the above methods reduce the impacts of different receivers, they require multiple labeled signal datasets collected by different receivers, leading to more costly expenses. Moreover, all of these methods are  built upon  intuitions and lack of theoretical support. 
		
		It is worth noting that the aforementioned methods primarily build upon domain alignment techniques originally developed for image. However, considering the cross-receiver RFFI problem, there are some specific challenges. Firstly, significant domain discrepancy where the impact of different receivers on signals far exceeds that seen in image classification, necessitating a more robust domain alignment strategy. Secondly, unlike in image classification, where the overall content of an image provides clear category indicators, RFFI requires a more robust pseudo-labeling strategy to generate reliable pseudo-labels due to the subtlety of the signal features and the potential for low accuracy of the pseudo-labels.
		
		In this paper, we aim at answering the above questions via linking the cross-receiver RFFI problem with the domain adaptation~\cite{ben2010theory,lee2023weight}. We first establish a theoretical upper bound on the  emitter identification  error probability when the adapted RFFI model is applied to Rx-2. Then, guided by this upper bound,  we propose a novel model adaptation  strategy for solving the cross-receiver RFFI problem, which consists of  two key ingredients, namely, domain alignment and pseudo-labeling. The former aims at learning a domain-invariant feature space, and the latter provides highly accurate pseudo-labels for adaptation by implicitly transferring the label knowledge from Rx-1 to Rx-2. This approach circumvents the need for manual labeling, reducing both time and resource costs. % Experiments conducted with real-world RF datasets corroborate the superior performance of the proposed method when compared to existing  methods. 
		In summary, our contributions are as follows:
		\begin{itemize}
			\item We develop a theoretical generalization error bound for the cross-receiver RFFI problem to analyze the performance of the model on the new receiver.
			\item  Motivated by the analysis, we propose a novel method to effectively reduce the impact of receivers   through domain alignment, pseudo-labeling,  and adversarial training. 
			%, thereby  reduce the impact of receiver and prove theoretically that the method converges well to a stationary point.
			\item Experiments conducted with real-world RF datasets corroborate the superior performance of the proposed method when compared to existing methods.
		\end{itemize}
	
        {\textit{Notation}: Throughout this paper, the superscript ``$s$'' and ``$t$'' indicate the source domain and the target domain, respectively. And the symbols $x$ and $y$ represent the received signal and its corresponding emitter label, respectively. $z$ represents the features obtained from the signal $x$ after passing through the model. $\mathbb{E}$ denotes the expectation, and $\mathds{1}_{\sf condition}$ is an indicator function with the value one if  the {\footnotesize \sf {condition}} is satisfied, and 0 otherwise. Additional notations are defined in Table~\ref{table:notaion}. }

		\begin{table*}[ht]
			\renewcommand\arraystretch{1.1}
			\centering
			
				\caption{\label{table:notaion} Notations used throughout the paper}
				\begin{tabular}{l l | l l}
					\hline
					Notation  & Definition & Notation & Definition \\
					\hline
					$\mathcal{D}^s$ & The distribution of signals from the source domain (Rx-1) & $\epsilon_\mu^{s\&t}(\cdot)$ & The weighted classification error probability on both domains \\
					$\mathcal{S}$ & The labeled dataset with Rx-1 & $h^\star$ & The optimal model on the target domain \\
					$\mathcal{D}^t$ & The distribution of signals from the target domain (Rx-2) & ${\sf E}(\cdot; \theta_{\sf E})$ & The mapping of the feature extractor network with parameters $\theta_{\sf E}$ \\
					$\mathcal{T}$ & The labeled dataset with Rx-2 & ${\sf C}(\cdot; \theta_{\sf C})$ & The mapping of the classifier network with parameters $\theta_{\sf C}$\\
					$f^t$ & The ground truth label function on the target domain & ${\sf T}(\cdot; \theta_{\sf T})$ & The mapping of the estimate network with parameters $\theta_{\sf T}$\\
					$\hat{f}^t$ & A pseudo-label function on the target domain & $\delta(\cdot)$ & The Softmax function \\	
					$\epsilon^t(\cdot, \cdot)$ & Probability that two functions have different results on $\mathcal{D}^t$ & $n^s$ & The number of training samples from $\mathcal{S}$ for one batch \\
					$\hat{\epsilon}^t(\cdot, \cdot)$ & Probability that two functions have different results on $\mathcal{T}$ & $n^t$ & The number of training samples from $\mathcal{T}$ for one batch \\
					\hline
			\end{tabular}
		\end{table*}

		%  from one where the training data and the test data . Herein, the 

		\section{System Model and Problem Statement}\label{sec:formulation}

		Let us start with the received signal model of  RFFI.  During the $\ell$-th time interval, the received signal  can be expressed as % \vspace{-5pt}
		\begin{equation} \label{eq:system_model}
			x_\ell(t) = \psi \left(c(t) * \varphi \left(s(t)\cos(\omega_0 t+\theta)\right)\right)+ n(t),
		\end{equation}
		where $(\ell-1)T \leq t\leq \ell T$, $n(t)$ is the noise, $T$ is the length of the interval, $\omega_0$ is the carrier frequency, $s(t)$ is random modulating signal,  $c(t)$ denotes the channel response,  the operator $*$ denotes convolution,  $\varphi(\cdot)$ is a random function modeling nonlinear distortion induced by the emitter's hardware, which encodes the ``fingerprint'' of  the emitter, and  $\psi(\cdot)$ is  a random function modeling the nonlinear reception characteristics induced by the receiver's hardware. Notice that different emitters and receivers have distinct  $\varphi$ and $\psi$, respectively (resp.). 
		
		\begin{figure}%[htb] 
			\begin{minipage}[b]{.48\linewidth}
				\centering
				\centerline{\includegraphics[width=3.8cm]{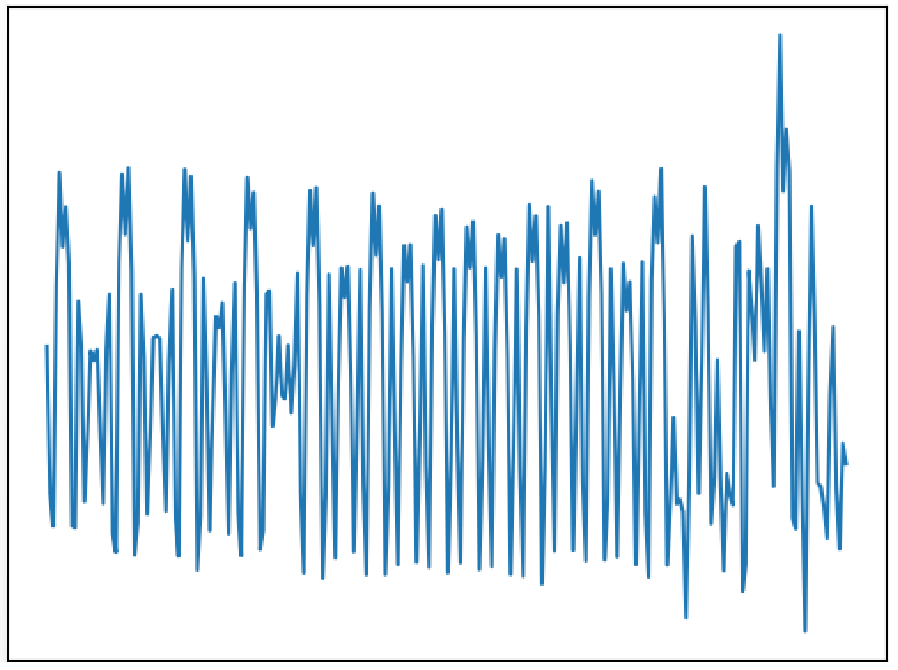}}
				%  \vspace{1.5cm}
				\centerline{(a) Waveform}\medskip
			\end{minipage}
			\hfill
			\begin{minipage}[b]{0.48\linewidth}
				\centering
				\centerline{\includegraphics[width=3.8cm]{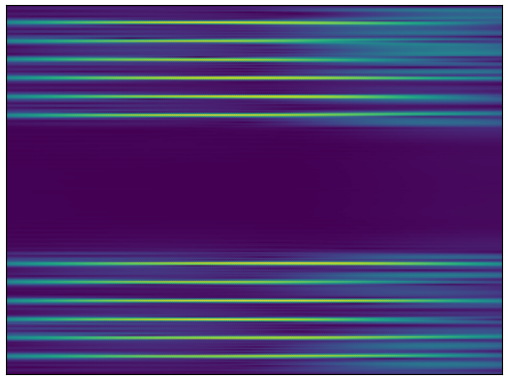}}
				%  \vspace{1.5cm}
				\centerline{(b) Spectrogram}\medskip
			\end{minipage}
			\caption{Waveform and spectrogram of the Wisig~\cite{hanna2022wisig} signals.}
			\label{fig:Spectrogram}     
		\end{figure}
	
		In this work, we use off-the-shelf Wisig dataset~\cite{hanna2022wisig}, which consists of signals from WiFi transmitters, to create the training and test data for RFFI. The specific data creation procedure involves  energy detection, channel equalization, and signal normalization.

		1) \textit{Energy Detection}: The signal detection   compares the energy of   $x_\ell(t)$  within  a period $T$ with  a threshold $\eta$ to determine whether there is a signal. If $\int_{T} |x_\ell(t)|^2 dt> \eta$, then $x_\ell(t)$ is recorded and passed to the subsequent processing; otherwise, $x_\ell(t)$ is dropped. 		
		%sub  the window exceeds the threshold, it is identified as signals. Subsequently, WiFi packets with sufficiently long signal lengths are selected.
		
		2) \textit{Channel Equalization}: To mitigate the impact of channel $c(t)$, autocorrelation is applied to the Legacy Short Training fields (L-STF) preamble for accurate packet start detection. Upon identifying the packets, the channel is estimated using the Legacy Long Training fields (L-LTF), and the signal is equalized by  minimum mean square error (MMSE) principle. 
		%While channel equalization can partially address the channel impact $c(t)$, it has limited impact on the receiver impact $\psi$.
		
		3) \textit{Signal Normalization}: To avoid overfitting and improve model stability, signal normalization is performed by subtracting the mean of the signal and dividing by the square root of the power. Fig.~\ref{fig:Spectrogram} shows the waveform and spectrogram of  $x_\ell(t)$ after the above processing.

		As mentioned in Section~\ref{sec:intro}, we assume that the collected signals by Rx-1 are well labeled, and for notation simplicity, we use $x_\ell^s$ to indicate the processed signal $x_\ell(t)$ and $y_\ell^s\in {\cal K} \triangleq \{1, \ldots, K\}$ to indicate the emitter index or label associated with $x_\ell^s$. Herein, following the convention in domain adaptation literature, we use  the superscript $s$ in $x_\ell^s$ to indicate that the signal is from the  source domain ${\cal D}^s$, i.e. Rx-1. Then, the labeled training dataset with Rx-1 is represented as $\mathcal{S}=\{(x_1^s,y_1^s),\cdots,(x_{N^s}^s, y_{N^s}^s)\}$, where $N_s$ is the size of the dataset.
		Similarly,  the unlabeled dataset with Rx-2  is represented as $\mathcal{T}=\{x_1^t,\cdots,x_{N^t}^t\}$, where the superscript $t$   indicates that the signal $x_\ell^t$ is from  the target domain ${\cal D}^t$, i.e. Rx-2. 
		
		\begin{figure}[]
			\centering
			\centerline{\includegraphics[width=8.7cm]{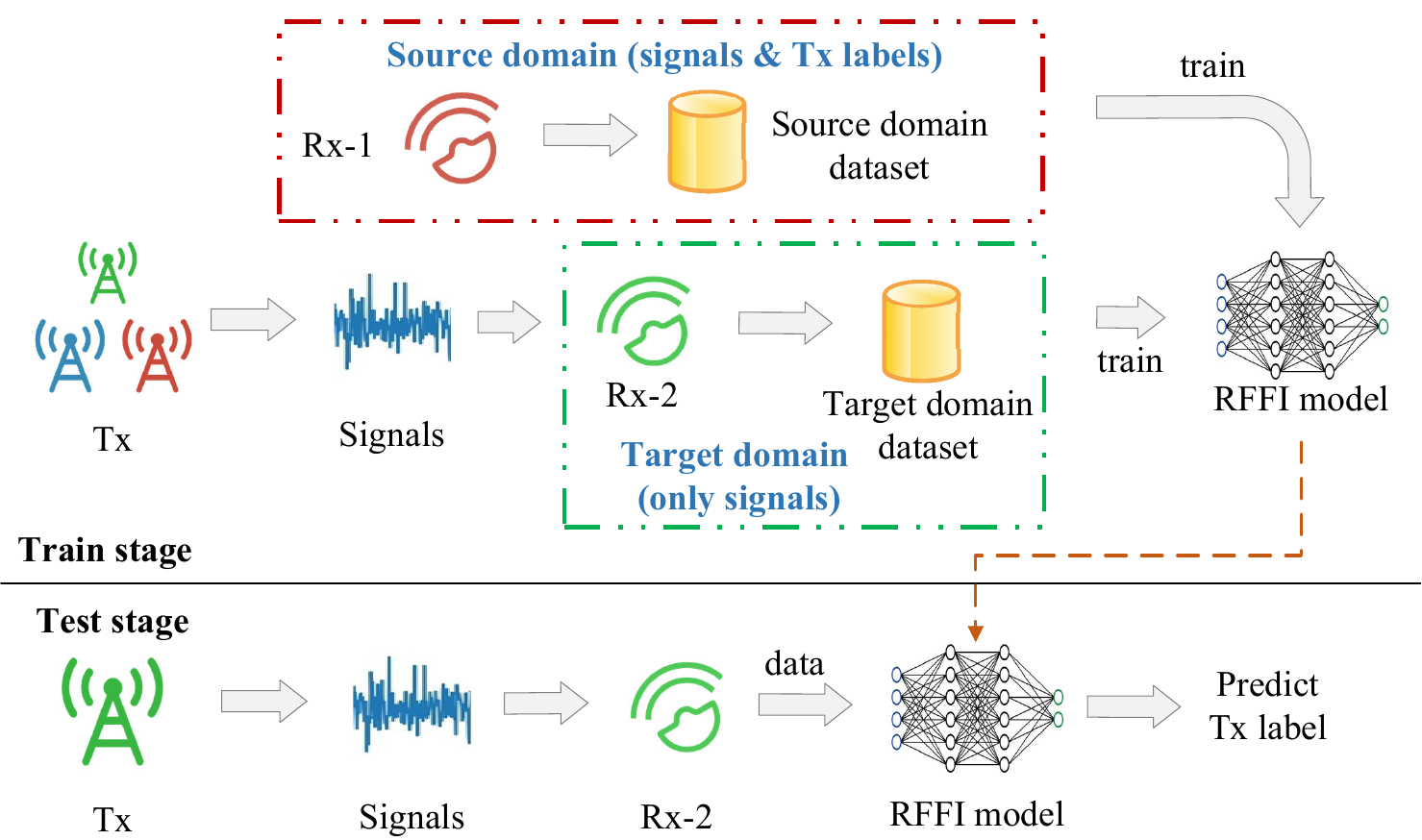}}
			%  \vspace{1.5cm}
			\caption{Scenario of cross-receiver RFFI.}
			\label{fig:scenario}
		\end{figure}
		Now, the cross-receiver RFFI problem can be described as follows (cf.~Fig.~\ref{fig:scenario}). Given the labeled dataset $\mathcal{S}$ from Rx-1 and the unlabeled data $\mathcal{T}$ from Rx-2, we seek for a classification model $h$ such that the classification error probability    on the target domain ${\cal D}^t$ is minimized, viz., %\vspace{-5pt}
		\begin{equation} \label{eq:goal}
			\begin{aligned}		
				& \min_{h \in {\cal H}}~\epsilon^t(h, f^t)  \triangleq {\mathbb{E}}_{X^t \sim \mathcal{D}^t}[ \mathds{1}_{h(X^t) \neq f^t(X^t)}  ] \\
				\approx~ & \min_{h \in \cal H} {\hat{{\epsilon}}^t (h, f^t)}  \triangleq \frac{1}{N^t}\sum_{i=1}^{N^t} \mathds{1}_{h(x_i^t) \neq y_i^t}  ,
			\end{aligned} 
		\end{equation}
		where $\epsilon^t(h, f^t)$ and ${\hat{{\epsilon}}^t (h, f^t)}$ denote expected risk and empirical risk on the target domain, resp., $\mathbb{E}$ denotes the expectation, taken over all the randomness in the received signal space, and  $f^t: {\cal D}^t\rightarrow \cal K$ denotes the (unknown) ground truth label function on the target domain, i.e. $f^t(x^t) = k$ if the signal $x^t$ is  from the emitter $k$, and  $\mathds{1}_{\sf condition}$ is an indicator function with the value one if  the {\footnotesize \sf {condition}} is satisfied, and 0 otherwise, and ${\cal H}$ is a presumed hypothesis space consisting of the learning models. We should mention that ${\cal H}$ is a functional space and its specific form depends on the structure and the size of the neural network. The  model's classification capability is determined by the complexity of  ${\cal H}$, usually measured by Vapnik-Chervonenkis (VC) dimension~\cite{vapnik1994measuring}; we will elaborate on this in the next section.
		
		Since we have no knowledge of $f^t$ or $y_i^t$, it is intractable to directly minimize $\epsilon^t(h, f^t) $ or ${\hat{{\epsilon}}^t (h, f^t)}$. In the ensuing sections, we will develop a tractable  surrogate for ${{{\epsilon}}^t (h, f^t)}$ to indirectly minimize $\epsilon^t(h, f^t) $.

		\section{Theoretical Analysis and Insights}
		Before delving into our proposed adaptation method, we will first answer the second question raised in Introduction by deriving an upper bound on the performance gap between the adapted model and the optimal model on ${\cal D}^t$. To simplify the analysis, let us consider the two emitters’ case, i.e. $\mathcal{K} = \{0, 1\}$ for binary classification. To this end, denote by $h^\star$  the optimal  model for problem~\eqref{eq:goal}. Since the objective $\epsilon^t(h, f^t)$ in~\eqref{eq:goal} depends on the unknown ground truth label function $f^t$ on ${\cal D}^t$, the $h^\star$ is generally not computable. To circumvent this difficulty, we resort to the pseudo-label technique and define  by $ \hat{f}^t: {\cal D}^t\rightarrow {\cal K}$   a pseudo-label function on the target domain. For example, the original learned RFFI model from Rx-1's training data may serve as a pseudo-label function for Rx-2. We will discuss how to learn a good $\hat{f}^t$ from $\cal S$ and $\cal T$ in the next section. Herein, for analysis, we just assume that there is an $\hat{f}^t$ available for model adaptation. Now, with ${\cal S}$, ${\cal T}$, and $\hat{f}^t$, a plausible adaptation formulation is to minimize the weighted classification error probability on both the source domain and the target domain, viz. %\vspace{-5pt}
		\begin{equation} \label{eq:joint_risk}
			\min_{h\in {\cal H}}~~\epsilon_\mu^{s\&t}(h) \triangleq  \mu \epsilon^s(h, f^s) + (1-\mu )  \epsilon^t(h, \hat{f}^t),% \vspace{-5pt}
		\end{equation}
		where the pseudo-label function $ \hat{f}^t$ is used for evaluating the error probability on the target domain, and $\mu \in (0,1)$ is a trade-off hyperparameter. The insight behind \eqref{eq:joint_risk} is that when both receivers have similar reception characteristics, i.e. ${\cal D}^s \approx {\cal D}^t$, it is expected that minimizing the error probability $\epsilon^s(h, f^s)$ on the source domain would also render small error probability on the target domain. Meanwhile, $\epsilon^t(h, \hat{f}^t)$ is a surrogate of the intractable $\epsilon^t(h, {f}^t)$. When $\hat{f}^t$ is close to the true label function $f^t$,  it is expected that  small $\epsilon^t(h, \hat{f}^t)$ should also imply small $\epsilon^t(h, {f}^t)$. The following theorem concretes the above insight.

		\begin{theorem}\label{theorem}
			Let $\hat{h}$ be an optimal model  for problem~\eqref{eq:joint_risk}.	Suppose that the hypothesis space $\mathcal{H}$ has a VC dimension $d$, and that $\mathcal{S}$ and $\mathcal{T}$ each has sample size $N$. Then, for any $\rho \in (0,1)$, with probability at least $1-\rho$ (over the choice of the samples),  the following inequality holds: %\vspace{-7pt}
			\begin{equation}\label{eq:theorem_bound}
				\begin{aligned}
					&	 \epsilon^t(\hat{h}, f^t) - \epsilon^t(h^\star, f^t)  \\
					\leq ~&   \mu d_{\mathcal{H}\Delta\mathcal{H}}(\mathcal{S}, \mathcal{T}) + 2 (1 + \mu) \epsilon^t(\hat{f}^t, f^t)  + 2 \mu \Lambda, \\
				\end{aligned}
			\end{equation}
			where $\epsilon^t(\hat{h}, f^t)$ and $\epsilon^t(h^\star, f^t)$ represent  the expected  error probability of $\hat{h}$ and   $h^\star$  on the target domain, resp.; $d_{\mathcal{H} \Delta \mathcal{H}}(\mathcal{S}, \mathcal{T}) = 2\sup_{h,h' \in \mathcal{H}}{ | \hat{\epsilon}^t(h, h') - \hat{\epsilon}^s(h, h')}|$ represents the domain discrepancy of $\mathcal{S}$ and $\mathcal{T}$ with respect to  $\mathcal{H}$,  $\Lambda = 4 \sqrt{\frac{2d\log{2N}+\log{\frac{2}{\rho}}}{N}} + \Lambda^\star$, and 
			$\Lambda^\star = \min_{h \in \mathcal{H}}{\epsilon^s(h, f^s) + \epsilon^t(h, f^t)}$ denotes the minimum combined error probability on both domains. 
		\end{theorem}
		
		The proof of Theorem~\ref{theorem} leverages on~\cite{ben2010theory} by taking into account the additional pseudo-labeling in the analysis. The detailed proof of Theorem~\ref{theorem} can be found in Appendix~\ref{appendix_proof_lemma}. Theorem~\ref{theorem} reveals that the performance gap depends on the domain discrepancy $d_{\mathcal{H} \Delta \mathcal{H}}(\mathcal{S}, \mathcal{T})$ and the  pseudo-label
		accuracy $\epsilon^t(\hat{f}^t, f^t)$. Specifically, when ${\cal D}^s \approx {\cal D}^t$ and $\hat{f}^t \approx f^t$,   the first two terms on the right-hand side of~\eqref{eq:theorem_bound} tend to zero, and consequently the performance gap is minimized.    
		Therefore, apart from minimizing $\epsilon_\mu^{s\& t}(h)$, it is also important to shrink the domain discrepancy as well as improve the pseudo-label accuracy during the  learning process.

		\section{Proposed Method}
		
		\begin{figure*}[]
			\centering
			\centerline{\includegraphics[scale=0.36]{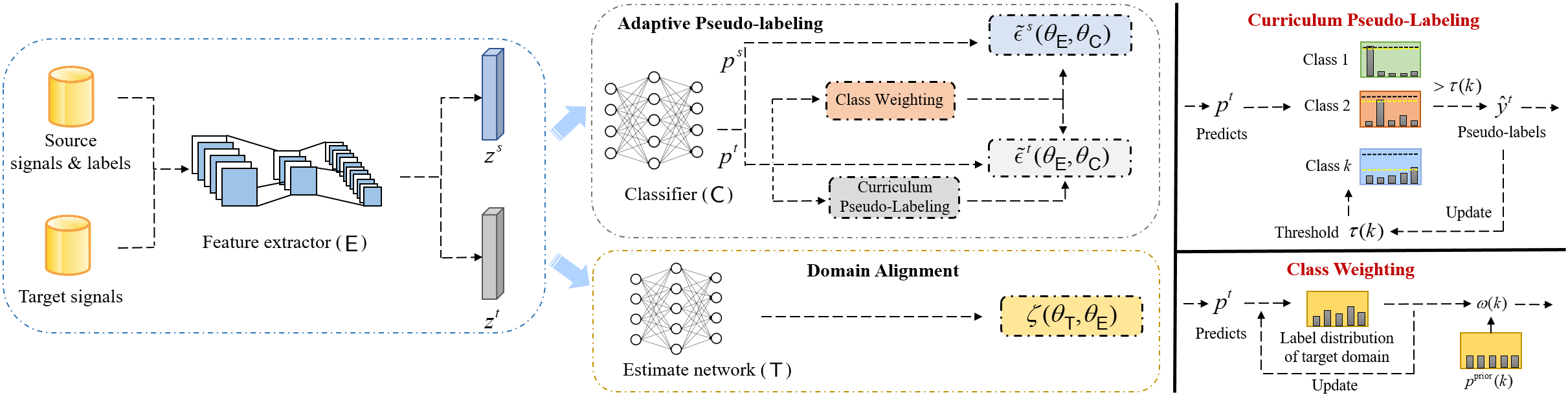}}
			%  \vspace{1.5cm}
			\caption{Overview of the proposed adaptation method. The proposed method involves domain alignment and adaptive pseudo-labeling techniques. First, the signals from the source and target domains are input into the feature extractor in order to obtain the signal features for the source and target domains, represented as $z^s$ and $z^t$ respectively. Then, on the one hand, domain alignment reduces the gap between the distributions of $z^s$ and $z^t$. On the other hand, adaptive pseudo-labeling ensures reliable, balance, and high-confidence pseudo-labeling.}
			\label{fig:overview}
		\end{figure*}
		
		\subsection{General Description of the Proposed Method}
		We start with an initial model $h^0$ (e.g. the learned model from Rx-1) and let $ \hat{f}^t=h^0 $. For every batch data, we update $h$ by  approximately  minimizing $\epsilon_\mu^{s\& t}(h)$ plus domain discrepancy. Then, the latest $h$ is used for updating $\hat{f}^t$ and generating pseudo labels for the next batch data. The above procedure repeats until the model $h$ converges.
		There are two key ingredients in our proposed method: domain alignment and adaptive pseudo-labeling. The former seeks for a feature space in which both domains have similar distributions, thereby shrinking the domain discrepancy,  and the latter introduces a  dynamic pseudo-labeling scheme to improve the pseudo-label  accuracy from batch to batch.

		Fig.~\ref{fig:overview} provides an overview of the proposed  adaptation method. It consists of three main components: the feature extractor network ($\sf E$), the classifier network ($\sf C$), and the  estimate network ($\sf T$). The estimate network ($\sf T$) is introduced to encourage domain alignment and will be dropped after training. Let $\theta_{\sf E}$, $\theta_{\sf C}$ and $\theta_{\sf T}$ be the respective  parameters of these networks. The learning of $h$ boils down to finding a set of network parameters $\theta_{\sf E}$, $\theta_{\sf C}$ and $\theta_{\sf T}$ such that  $\delta \left({\sf C} \left({\sf E}(x^t;{\theta_{\sf E}}); {\theta_{\sf C}} \right) \right) $  is close to the true label of $x^t$, where $\delta(\cdot)$ is the Softmax function,  ${\sf E}(\cdot;{\theta_{\sf E}})$  and ${\sf C}(\cdot;{\theta_{\sf C}})$ denote mapping of the  feature extractor network and   the classifier network, resp. In the following, we will formalize the idea of domain alignment and adaptive pseudo-labeling and develop an adversarial training algorithm to update  $\theta_{\sf E}$, $\theta_{\sf C}$ and $\theta_{\sf T}$.

		\subsection{Domain Alignment}
		Let $z^s={\sf E}(x^s; \theta_{\sf E})\in \mathbb{R}^{d_z}$ and  $z^t={\sf E}(x^t; \theta_{\sf E}) \in \mathbb{R}^{d_z}$ be the extracted features in the  source domain and the target domain, resp., where $d_z$ is the dimension of the feature space and $\sf E$ denotes the feature mapping. To achieve domain alignment, we propose to minimize the  
		Kullback-Leibler (KL) divergence between  $p(z^s)$ and $p(z^t)$, where $p(z)$ represents the distribution of the variable $z$. Since it is  hard to directly minimize KL divergence for neural networks, we exploit a variational form of KL divergence, also known as Donsker-Varadhan (DV) representation~\cite{donsker1975asymptotic}: %\vspace{-5pt}% of KL divergence: 
		\begin{equation}\label{eq:dual_KL}
			D_{KL}(p(z^s) || p(z^t)) = \mathop{\sup}_{T:\Omega \rightarrow \mathbb{R}} {\mathbb{E}_{p(z^s)}[T] - \log {\mathbb{E}_{p(z^t)}[e^T]}}, %\vspace{-5pt}
		\end{equation}
		to facilitate the minimization. In~\eqref{eq:dual_KL}, the supremum is taken with all continuous functions. %Inspired by \cite{belghazi2018mine}, w
		We approximate $T$ by a neural network ${\sf T}(\cdot;\theta_{\sf T}):\mathbb{R}^{d_z} \rightarrow \mathbb{R}$ with $\theta_{\sf T}$ being the network parameters, and replace $p(z^s)$ and $p(z^t)$ with their sample distribution over the 
		datasets ${\cal S}$ and ${\cal T}$, resp. Then, the domain alignment is achieved by solving the following  min-max optimization problem: %\vspace{-5pt}
		\begin{equation}\label{eq:min-max}
			\min_{\theta_{\sf E}}\max_{\theta_{\sf T}} ~ { \zeta(\theta_{\sf T},\theta_{\sf E})}, %\vspace{-5pt}
		\end{equation}
		where 
		\begin{equation*}
			{ \zeta(\theta_{\sf T},\theta_{\sf E})} \triangleq \frac{1}{n^s} \sum_{i=1}^{n^s}{{\sf T}({\sf E}(x_i^s))} - \log { \left [\frac{1}{n^t} \sum_{j=1}^{n^t}{e^{{\sf T}({\sf E}(x_j^t))}} \right ]},
		\end{equation*} 
		and $n^s$ and $n^t$ denote the number of training samples for one batch.

		\subsection{Adaptive Pseudo-Labeling}
		
		Let $\theta_{\sf E}^{\ell}$, $\theta_{\sf C}^{\ell}$ and $\theta_{\sf T}^{\ell}$ be the  network parameters after  the  $(\ell-1)$-th batch. A simple  pseudo-labeling scheme is to set a confidence parameter $\tau \in (0.5,~1)$  and let $\hat{y}_i^t = k$ if the $k$-th entry of the network output probability vector $\delta \left({\sf C} \left({\sf E}(x_i^t;{\theta_{\sf E}^{\ell}}); {\theta_{\sf C}^{\ell}} \right) \right)$ exceeds $\tau$, otherwise $x_i^t$ is not pseudo-labeled. However, the drawback with the  fixed $\tau$ is also clear---it is hard to set a universally  good $\tau$ for different emitters,  consequently leading to unbalanced pseudo-labels among emitters. To address the unbalanced pseudo-label problem, two strategies are employed: Curriculum Pseudo-Labeling (CPL)~\cite{zhang2021flexmatch} and class weighting.

		\subsubsection{Curriculum Pseudo-Labeling}
		CPL dynamically adjusts the threshold for each class based on the number of pseudo-labeled samples. A higher threshold should be set for classes with more pseudo-labeled samples. We advocate the following scaling strategy for class $k\in \cal K$ at the $\ell$-th batch: %\vspace{-5pt}
		%CPL adjusts the threshold for each class dynamically according to the number of pseudo-labeled samples of that class. The more samples of a class are pseudo-labeled,  the higher the threshold of this class should be. Therefore, the fixed threshold $\tau$ should be scaled  according to the respective learning status of each class. Herein, we advocate the following scaling strategy for class $c\in \cal K$:
		\begin{equation}\label{eq:beta} 
			\beta_\ell(k) = \frac{\sigma_{\ell-1}(k)}{\mathop{\max}_{i \in \cal K} {\sigma_{\ell-1}(i)}} \in [0,~1], 
		\end{equation}
		\begin{equation}\label{threshold}
			\tau_\ell(k) = \beta_{\ell}(k) \cdot \tau, 
		\end{equation}
		where $\sigma_{\ell-1}(k)$ denotes the number of accumulated pseudo-labeled samples of class $k$ until $(\ell-1)$-th batch, and $\tau_\ell(k)$ is the scaled threshold of class $k$ at $\ell$-th batch. We set by default $\beta_1(k)=1$ and $\tau_1(k) = \tau$ for all the emitters.

		\subsubsection{Class Weighting}
		Class weighting addresses the  unbalanced pseudo-label problem by considering  weighted classification error probability for different classes. Let $p^{\rm prior}(k)$ be prior classification probability for emitter $k$, which may be counted from Rx-1's the labeled data  ${\cal S}$.  If the empirical class distribution counted on the target domain for class $k$ is greater than its prior probability $p^{\rm prior}(k)$, the weight of class $k$ in the classification loss should be lowered, and vice versa. In this way, the class weight for class $k$ at $\ell$-th batch is proposed as %\vspace{-5pt}
		\begin{equation}
			\label{class_weight}
			\omega_\ell(k) =  \frac{p^{\rm prior}(k)}{ \sigma'_{\ell-1}(k) / n_{\ell-1}^t}, 
		\end{equation}
		where  $n_{\ell-1}^t$ is the total number of samples in the target domain up to $(\ell-1)$-th batch, $\sigma'_{\ell-1}(k)$ denotes the  number of accumulated samples that have been classified as class $k$ up to $(\ell-1)$-th batch  , and thus $\sigma'_{\ell-1}(k) / n_{\ell-1}^t$ can be seen as an estimate of $p^{\rm prior}(k)$.

		With CPL and  class weighting, the weighted classification error probability in~\eqref{eq:joint_risk} is modified as %\vspace{-5pt}
		\begin{equation}\label{eq:CE}
			\tilde{\epsilon}_\mu^{s\&t}(\theta_{\sf E}, \theta_{\sf C} ) \triangleq \mu \tilde{\epsilon}^s  + (1-\mu )  \tilde{\epsilon}^t,
			%\vspace{-5pt}
		\end{equation}
		where  %\vspace{-5pt}
		\begin{equation*}
			\tilde{\epsilon}^s   \triangleq \frac{1}{n^s}\sum_{i=1}^{n^s} \omega_\ell(y_i^s)  {\mathsf H} \big(p_i^s, y_i^s \big),
		\end{equation*}
		\begin{equation*}
			\tilde{\epsilon}^t  \triangleq \frac{1}{|{\mathcal{T}}^{\ell}|} \sum_{(x_i^t, \hat{y}_i^t) \in {\mathcal{T}}^{\ell}} \omega_\ell(\hat{y}_i^t) {\mathsf H} \big(p_i^t, \hat{y}_i^t \big),
		\end{equation*}
		are sample average approximation to the expected error probability on the source and the domain, resp., and   ${\cal T}^{\ell} \in {\cal T}$ be the pseudo-labeled data within $\ell$-th batch, $ {\mathsf H} \big(p_i^s, y_i^s \big)$ denotes the cross-entropy between $p_i^s$ and $y_i^s$ with $p_i^s = \delta \left({\sf C} \left({\sf E}(x_i^s;{\theta_{\sf E}}); {\theta_{\sf C}} \right) \right)$ being the softmax output of the network.  Similar notations apply for ${\mathsf H} \big(p_i^t, \hat{y}_i^t \big) $ and $p_i^t$.

		\begin{algorithm}[t]
			\caption{\small Gradient ascent-descent (GAD) adversarial training algorithm of the proposed method}
			\label{method}
			\textbf{Input:} 
			\begin{algorithmic}
				\State $\mathcal{S}$: Labled source domain data $\{{(x_i^s,y_i^s)}\}_{i=1}^{N^s}$
				\State $\mathcal{T}$: Unlabled source domain data $\{{(x_i^t)}\}_{i=1}^{N^t}$
				\State ${\sf E}, \theta_{\sf E}$: Feature extractor and its parameters
				\State ${\sf T}, \theta_{\sf T}$: Estimate network and its parameters
				\State ${\sf C}, \theta_{\sf C}$: Classifier and its parameters
				\State $m$: Update frequency of estimate netwrok
				\State $\tau$: Preset threshold of pseudo-label
				\State $b$: Batch size
				\State $\lambda, \eta$: Hyper-parameter
			\end{algorithmic}
			%	\textbf{Initialize:} $k=0$, ${X}^k$, and $w^k_i$, $i=1,\cdots,m$.\\
			\textbf{Output:}
			\begin{algorithmic}
				\State $\theta_{\sf E}, \theta_{\sf C}$: Network parameters
			\end{algorithmic}
			\begin{algorithmic}[1]
				\While {not convergent}
				\State {$\sigma_0(k), \sigma'_0(k), n_0^t \gets 0$}
				\For {$\ell = 1, 2, \cdots, \min{\left\{ N^s/b, N^t/b) \right\}}$}
				\State Get data of batch $\ell$ from $\mathcal{S} \cup \mathcal{T}$: $x^s, y^s, x^t$
				\For {$i = 1, 2, \cdots, m$}
				\State Compute $\zeta(\theta_{\sf T},\theta_{\sf E})$ in Eqn.~\eqref{eq:min-max}
				\State $\theta_{\sf T} \gets \theta_{\sf T} + \eta_{\sf T,\ell} \cdot\frac{\partial \zeta}{\theta_{\sf T}}$
				\EndFor
				%		\State Compute $\mathcal{L}_{KL}$ by Eq. \eqref{kld}
				\State Compute threshold $\tau_\ell(k)$ by Eqn. \eqref{threshold}
				\State Pseudo-label samples $x^t$ and get sub-dataset $\mathcal{T}^\ell$
				\State Compute class weight $\omega_\ell(k)$ by Eqn. \eqref{class_weight}
				\State Compute  the weighted loss $\cal L$ by Eqn.~\eqref{eq:loss_all_eqv}
				%				\State Form ${\cal L}= \mathcal{L}_{CE}^s + \mathcal{L}_{CE}^t + \lambda \mathcal{L}_{KL}$
				\State $\theta_{\sf E} \gets \theta_{\sf E} - \eta_{\sf E,\ell} \cdot\frac{\partial {\cal L}}{\theta_{\sf E}}$
				\State $\theta_{\sf C} \gets \theta_{\sf C} - \eta_{\sf C,\ell} \cdot\frac{\partial {\cal L}}{\theta_{\sf C}}$
				\State Update $\sigma_\ell(k), \sigma'_\ell(k), n_\ell^t$
				\EndFor
				\EndWhile
				\State \Return $\theta_{\sf E}, \theta_{\sf C}$
			\end{algorithmic}
		\end{algorithm}
		
		\subsection{Training}
		Combining \eqref{eq:min-max} and\eqref{eq:CE}, we arrive at the loss function that we intend to optimize %\vspace{-5pt}
		\begin{equation}\label{eq:loss_all_eqv}
			\mathop{\min}_{\theta_{\sf E}, \theta_{\sf C}}  {\mathop{\max}_{\theta_{\sf T}}~ {\cal L}(\theta_{\sf E}, \theta_{\sf C},\theta_{\sf T}) \triangleq \tilde{\epsilon}_\mu^{s\&t}(\theta_{\sf E}, \theta_{\sf C} )  + \lambda\zeta(\theta_{\sf T},\theta_{\sf E}) }, %\vspace{-5pt}
		\end{equation}
		where $\lambda > 0$ is a hyperparameter for balancing domain alignment and pseudo-labeling. Problem~\eqref{eq:loss_all_eqv} is a  minmax problem. It can be adversarially trained by alternately performing gradient ascent with  respect to $\theta_{\sf T}$ and gradient descent with  respect to $(\theta_{\sf E}, \theta_{\sf C})$ until some stopping criterion is satisfied. The Gradient Ascent-Decent (GAD) adversarial training algorithm of problem~\eqref{eq:loss_all_eqv} is shown in Algorithm~\ref{method}.
		Since problem~\eqref{eq:loss_all_eqv} is non-convex, Algorithm~\ref{method} may fail to converge. Nevertheless, the  following result identifies the convergence of the GAD for linear networks.

		\begin{theorem}\label{theorem:2}
			Assume that  $\sf T$, $\sf E$ and $\sf C$ are linear networks and that $\theta_{\sf T}$, $\theta_{\sf E}$ and $\theta_{\sf C}$ are  within some convex and compact sets. Then,
			with appropriate  chosen step size of the GAD,  the adversarial training in algorithm~\ref{method}  converges to an $\epsilon$-stationary point of the minimax problem~\eqref{eq:loss_all_eqv}. 
		\end{theorem}
		The key to the proof of Theorem~\ref{theorem:2} is to exploit the linear structure of the network as well as the special form of the loss function in~\eqref{eq:loss_all_eqv} to show that the GAD training satisfies  the  convergence conditions  in~\cite[Theorem 4.8]{lin2020gradient}. The detailed proof can be found in Appendix~\ref{appendix_proof_claim}.

		\section{Experiment} \label{sec:experiment}

		\subsection{Setups}

		In this section, we evaluate our method using the Wisig dataset~\cite{hanna2022wisig}. We  pick  a subset of the Wisig dataset with  six transmitters (Tx) and 12 receivers (Rx) and the signals are collected over four days. The received signals undergo energy detection, channel equalization, and signal normalization, as mentioned in Section \ref{sec:formulation}, to form the dataset. The dataset is divided into 12 domains, each corresponding to one receiver's data collected over four days. We perform cross-receiver tests, randomly selecting two receivers as source and target domains, resulting in four cross-receiver tasks as shown in Table~\ref{results_other_methods}. As the receiver is positioned within a two-dimensional (2D) plane, we use two-dimensional coordinates ``$x-y$'' for distinct receiver identification. And ``$x-y \rightarrow x'-y'$'' means adapting from signals collected by Rx ``$x-y$'' to signals collected by Rx ``$x'-y'$''. 
		
		In addition, we also perform the cross-day test to test the performance of the proposed method under  temporal domain variation induced by channel or environment change. Then the dataset is divided into 4 domains, each of which corresponds to one day. We choose signals collected on 1st March and 23rd March. The ``d01 $\rightarrow$ d23'' task in Table~\ref{results_other_methods} means adapting from signals collected on 1st March to signals collected on 23rd March.

		The feature extractor uses ResNet18~\cite{he2016deep}. We replace the 2D-convolution with 1D-convolution in ResNet18 to better accommodate the one-dimensional time series. The classifier and estimation networks are implemented as three-layer fully connected networks. Hyperparameters include a learning rate of $0.0006$, a KL divergence loss weight of $\lambda = 0.005$, the trade-off hyperparameter $\mu=0.5$, the update frequency of the estimate network $m=7$, and an initial pseudo-label threshold of $\tau = 0.7$.

		\begin{table*}[ht]
			\renewcommand\arraystretch{1.3}
			\centering
			\caption{\label{results_other_methods}Classification accuracies (\%) comparison with other methods}
			\begin{tabular}{l c c c c c}
				\hline
				Method  & d01 $\rightarrow$ d23 & 14-7 $\rightarrow$ 3-19 & 1-1 $\rightarrow$ 1-19 & 1-1 $\rightarrow$ 8-8 & 7-7 $\rightarrow$ 8-8 \\
				\hline
				Souce only & 83.69 $\pm$ 0.19 & 30.25 $\pm$ 0.48 & 67.87 $\pm$ 1.04 & 59.94 $\pm$ 0.49 & 55.50 $\pm$ 0.80\\
				DANN \cite{ganin2016domain}  & 90.48 $\pm$ 0.07 & 58.16 $\pm$ 4.73 & 74.82 $\pm$ 0.10 & 96.52 $\pm$ 0.81 & 70.32 $\pm$ 0.35\\
				MCD \cite{saito2018maximum} & 87.90 $\pm$ 0.01 & 66.24 $\pm$ 0.10 & 79.64 $\pm$ 0.03 & 62.75 $\pm$ 0.01 & 66.95 $\pm$ 0.14\\
				SHOT \cite{liang2020we} & 86.04 $\pm$ 0.04 & 80.45 $\pm$ 0.02 & 79.37 $\pm$ 0.02 & 83.21 $\pm$ 0.00 & 66.72 $\pm$ 0.01\\
				\hline
				Proposed & \textbf{93.34 $\pm$ 0.02} & \textbf{92.42 $\pm$ 0.16} & \textbf{95.44 $\pm$ 0.51} & \textbf{99.78 $\pm$ 0.01} & \textbf{99.74 $\pm$ 0.04} \\
				
				\hline
			\end{tabular}
		\end{table*}
		
		\subsection{Comparison with Existing Methods}

		We compare the proposed method with existing methods: DANN~\cite{ganin2016domain}, MCD~\cite{saito2018maximum}, and SHOT~\cite{liang2020we}. In addition, as a benchmark, we will also present the results of the source only training method, which uses only the source domain signals to train a classification model.
		
		Table~\ref{results_other_methods} shows the classification accuracy of the tasks. As seen, all the methods outperform the source only model, which does not consider the cross-receiver effect on RFFI. Overall, the proposed method outperforms the other  methods for all of the tasks.
		As for the cross-receiver tests shown in the last four columns in Table~\ref{results_other_methods}, the proposed method significantly outperforms the other methods in most tests.  In particular,  the classification accuracies of the proposed method in  ``1-1 $\rightarrow$ 8-8'' and ``7-7 $\rightarrow$ 8-8'' tasks are close to 100\%.
		
		Notice that the proposed method performs the best for the challenging tasks, named ``d01 $\rightarrow$ d23''. This demonstrates that while the method is originally proposed to solve the cross-receiver RFFI problem, it also has great potential in achieving environmental-independent RFFI, which is another challenging problem arising from varied environments/channels during training and testing.
		
		In our analysis of the experimental results, particularly the significant differences in classification accuracy for tasks like 14-7 $\rightarrow$ 3-19 and 1-1 $\rightarrow$ 1-19, we attribute these variations primarily to the unique characteristics of different receivers' radio frequency fingerprints. We believe that variations in data distribution across different receivers, each with their unique signal signature, significantly influence the classification outcomes. The closer the similarity in fingerprints between two receivers, the more consistent the signal distribution they have, leading to higher accuracy when a model trained on one receiver is applied to another. 
		
		For a deeper analysis, our method to cross-receiver RFFI adaptation is closely related to ``Adversarial Domain Adaptation'', such as DANN. The proposed method introduces a unique network structure instead of the domaoin classifier. This structure incorporates an estimation network dedicated to domain alignment based on KL divergence. Alongside this, we employ a customized loss function that merges weighted classification error with domain discrepancy. Additionally, our method integrates an adaptive pseudo-labeling technique, which significantly improves the use of unlabeled data from the target domain, thus expediting the adaptation process of the model. It is precisely these enhancements that result in the proposed method's performance (the last row) being significantly superior to DANN (second row).

		\begin{table*}[ht]
			\renewcommand\arraystretch{1.3}
			\centering
			\caption{\label{results_ablation}Classification accuracies (\%) of experiments for ablation studies}
			\begin{tabular}{c c c c c c c c}
				\hline
				\multicolumn{3}{c}{Components} & \multirow{2}{*}{d01 $\rightarrow$ d23} & \multirow{2}{*}{14-7 $\rightarrow$ 3-19} & \multirow{2}{*}{1-1 $\rightarrow$ 1-19} & \multirow{2}{*}{1-1 $\rightarrow$ 8-8} & \multirow{2}{*}{7-7 $\rightarrow$ 8-8}\\
				\cline{1-3}
				Domain Alignment & CPL & Class Weighting & ~ & ~ & ~ & ~ & ~ \\
				\hline
				$\times$ & $\times$ & $\times$ & 83.69 $\pm$ 0.19 & 30.25 $\pm$ 0.48 & 67.87 $\pm$ 1.04 & 59.94 $\pm$ 0.49 & 55.50 $\pm$ 0.80 \\
				\hline
				\checkmark & $\times$ & $\times$ & \textbf{93.44 $\pm$ 0.02} & 76.36 $\pm$ 0.29 & 76.62 $\pm$ 0.07 & 99.69 $\pm$ 0.03 & 99.73 $\pm$ 0.14 \\
				$\times$ & \checkmark & $\times$ & 91.13 $\pm$ 0.02 & 54.68 $\pm$ 0.02 & 83.26 $\pm$ 0.00 & 55.07 $\pm$ 0.03 & 50.50 $\pm$ 0.01 \\
				$\times$ & $\times$ & \checkmark & 85.28 $\pm$ 0.17 & 35.83 $\pm$ 0.36 & 75.28 $\pm$ 0.40 & 59.90 $\pm$ 0.65 & 45.77 $\pm$ 1.50 \\
				\hline
				\checkmark & \checkmark & $\times$ & 91.30 $\pm$ 0.02 & 59.48 $\pm$ 0.04 & 82.84 $\pm$ 0.01 & 57.55 $\pm$ 0.03 & 54.50 $\pm$ 0.03 \\
				\checkmark & $\times$ & \checkmark & 92.11 $\pm$ 0.02 & 77.02 $\pm$ 0.19 & 76.67 $\pm$ 0.07 & 99.71 $\pm$ 0.02 & 99.64 $\pm$ 0.12 \\
				$\times$ & \checkmark & \checkmark & 92.77 $\pm$ 0.01 & 77.11 $\pm$ 0.17 & \textbf{99.17 $\pm$ 0.04} & 96.05 $\pm$ 0.01 & 98.35 $\pm$ 0.52 \\
				\hline
				\checkmark & \checkmark & \checkmark & 93.34 $\pm$ 0.02 & \textbf{92.42 $\pm$ 0.16} & 95.44 $\pm$ 0.51 & \textbf{99.78 $\pm$ 0.01} & \textbf{99.74 $\pm$ 0.04} \\
				
				\hline
			\end{tabular}
		\end{table*}
	
		\begin{figure}
			\centering
			\includegraphics[scale=0.5]{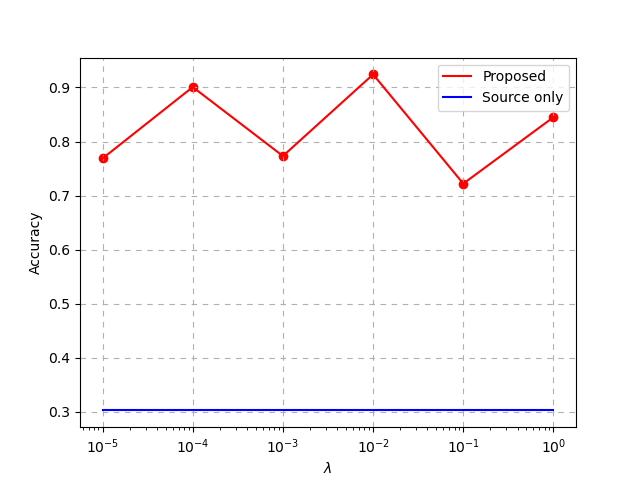}
			\caption{The curve of target receiver data classification accuracy with weight of estimated KL divergence $\lambda$.}
			\label{lambda}
		\end{figure}
		
		\subsection{Ablation Studies}
		We do ablation studies on each component of the proposed method. The results are shown in Table~\ref{results_ablation} and demonstrate the effectiveness of each component in improving classification accuracy.
		
		As shown in Table~\ref{results_ablation}, domain alignment can significantly improve the performance on the target domain. More specifically, with domain alignment (second row),  the accuracy is improved by 
		10\% as compared with the source only training (first row), which demonstrates the effectiveness of domain alignment. In addition, domain alignment also performs well with other components. For example, the proposed method (the last row) outperforms the method without domain alignment (penultimate row) in most cases.
		
		And only training with CPL (third row) sometimes boosts the classification accuracy compared to the source only training (first row), whereas sometimes worsens the accuracy. The reason for this is that even with adaptive thresholding,  there is still a probability to incorrectly pseudo-label the target signal,  and this probability is related to the discrepancy of the two receivers' impacts. 
		
		When applying only the class weighting technique (fourth row), compared with source only (first row), there is little influence on the results. However, when class weighting is accompanied by CPL, it achieves notable improvement in classification accuracy and  even outperforms the proposed method in some tests, e.g. the ``1-1 $\rightarrow$ 1-19'' task. We believe the reason is that the proposed model solves a challenging minimax problem, and for some hard cases, the training of the proposed model may get stuck at some bad stationary point. 
		
		Moreover, joint use of the three components (domain alignment, CPL, and class weighting) in the proposed method  is beneficial in general, which also illustrates the effectiveness of each component.

		\subsection{Hyper-parameter Sensitivity}
		In this subsection, we study the hyper-parameter sensitivity for the proposed method. We consider the ``14-7 $\rightarrow$ 3-19'' task for illustration, and the results for other tests are similar.

		\subsubsection{Sensitivity of the weight $\lambda$ for domain alignment}

		To showcase the sensitivity of the classification accuracy with respect to the weight $\lambda$, we train the proposed method with different $\lambda$. The results are shown in Fig.~\ref{lambda}. As seen,   the classification accuracy on receiver ``3-19'' varies slightly as the weight $\lambda$ increases from $10^{-5}$ to 1, and  substantially higher than that of the source only model. This demonstrates that the proposed method is robust against  $\lambda$. However, it should be noted that high value of $\lambda$ does not imply high classification accuracy. Therefore, it is needed to balance the domain alignment and the pseudo-labeling in order to maximize the classification accuracy.
		
		\subsubsection{Sensitivity of pre-set pseudo-label threshold $\tau$} 
		
		\begin{figure}
			\centering
			\includegraphics[scale=0.5]{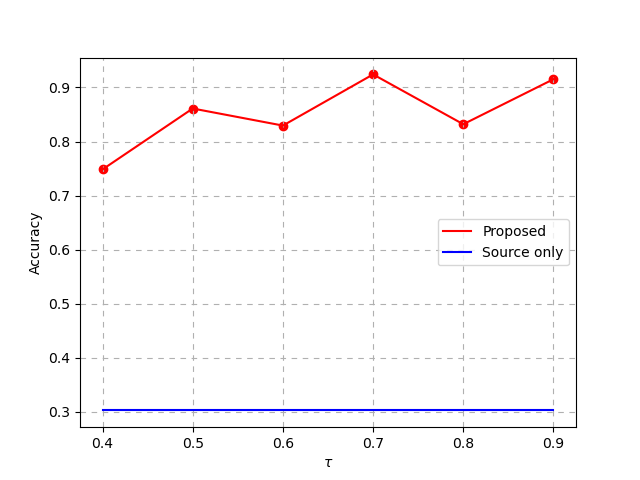}
			\caption{The curve of target receiver data classification accuracy with pre-set threshold of pseudo-label $\tau$.}
			\label{tau}
		\end{figure}
		
		Next, let us investigate the sensitivity of classification accuracy concerning the pre-set pseudo-labeling threshold $\tau$. The result is shown in Fig.~\ref{tau}. As seen from the figure, the classification accuracy  is overall worsened for a small threshold $\tau$. This is because for small $\tau$,  more samples in the target domain tend to be pseudo-labeled at the initial training stage, but since the model is far from convergence, lots of the target domain samples may be mis-labeled and lead to error propagation for the subsequent training.

		\begin{table}[ht]
			\renewcommand\arraystretch{1.3}
			\centering
			\caption{\label{results_prior}Classification accuracies (\%) of experiments for different estimated methods of prior probability of target domain labels (``$x$ CH'' means the numbers of training samples for $x$ number of classes are halved. Specifically, The notation "0CH" indicates that the number of training samples for all classes remains unchanged compared to previous experiments.).}
			\begin{tabular}{l c c c c c}
				\hline
				Method & 0 CH  & 1 CH & 2 CH  & 3 CH & Avg.\\
				\hline
				No class weighting & 59.48 & 58.57 & 58.28 & 46.65 & 55.75 \\
				${\rm Uniform}$ & \textbf{92.42} & 55.52 & \textbf{88.95} & 65.57 & 75.62 \\
				$p^{\rm prior}(k)$ & \textbf{92.42} & \textbf{88.48} & 83.21 & \textbf{88.10} & \textbf{88.05} \\
				
				\hline
			\end{tabular}
		\end{table}	
		
		\subsubsection{Sensitivity of the  prior probability ${p}^{\rm prior}(k)$}
		In all the previous experiments, we have assumed ${p}^{\rm prior}(k)$ is uniform and exactly known during training when using class weighting. However, in practice,  ${p}^{\rm prior}(k)$  may be non-uniform and also unknown. To simulate this case, herein we intentionally create non-uniform ${p}^{\rm prior}(k)$ by reducing the number of training samples for some classes by half. With more classes' training data halved, ${p}^{\rm prior}(k)$ tends to be more non-uniform.  We want to investigate how robust the proposed method is against the mismatch between the  presumed uniform $p^{\rm prior}(k)$ in training and the true non-uniform $p^{\rm prior}(k)$. The results are shown in Table~\ref{results_prior}. For comparison, we also list the result of no class weighting, i.e., $\omega_\ell(c)=1~\forall~\ell, c$, and the result with   true ${p}^{\rm prior}(k)$.   As shown in the table, the method with knowledge of ${p}^{\rm prior}(k)$ performs the best on  average, and then followed by the uniform distribution-based weighting method. The accuracy of no weighting is far inferior to the other class weighting methods. 
		From this experiment, we see that while there is a mismatch between the presumed uniform $p^{\rm prior}(k)$ and the true non-uniform $p^{\rm prior}(k)$, the uniform distribution-based weighting strategy is still able to achieve significant improvement over the no weighting one. 
		
		\subsection{t-SNE Visualization}
\begin{figure}[htb] 
			\begin{minipage}[b]{.49\linewidth}
				\centering
				\centerline{\includegraphics[width=4.4cm]{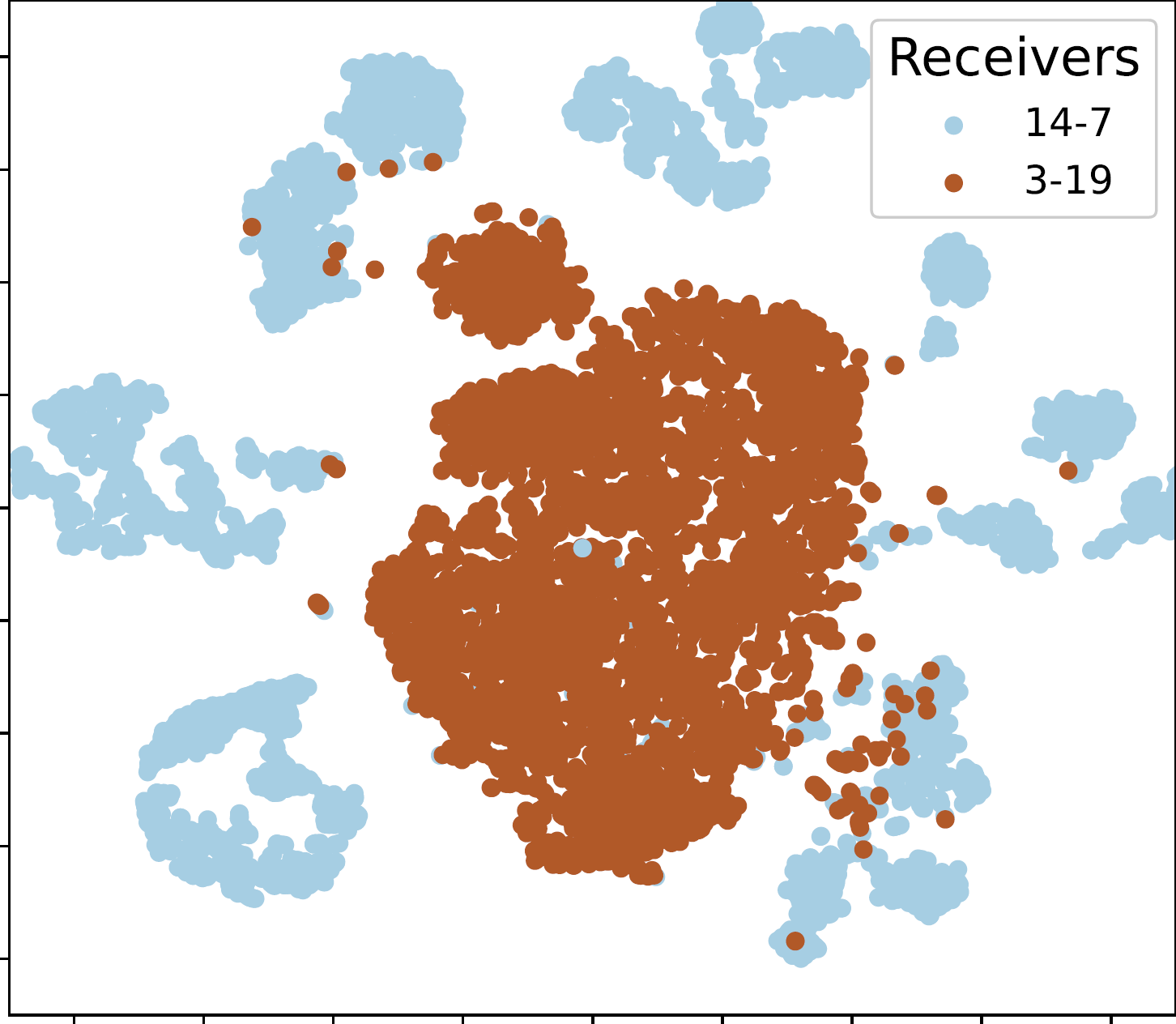}}
				%  \vspace{1.5cm}
				\centerline{(a) Source only}\medskip
			\end{minipage}
			\hfill
			\begin{minipage}[b]{0.49\linewidth}
				\centering
				\centerline{\includegraphics[width=4.4cm]{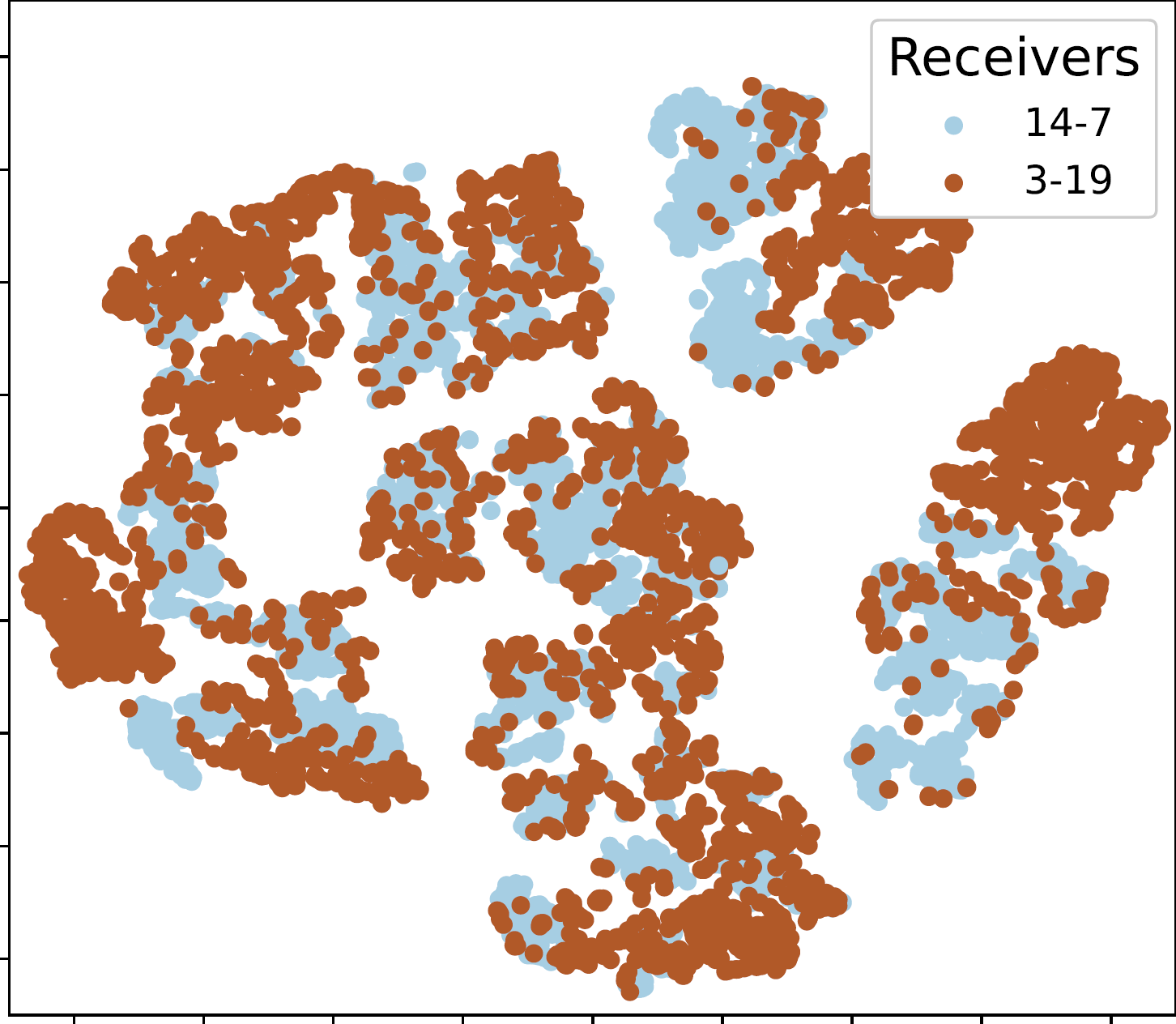}}
				%  \vspace{1.5cm}
				\centerline{(b) Proposed method}\medskip
			\end{minipage}
			\caption{Visualization of source/target feature distribution on (a) Source only,  (b) Proposed method (blue: source signals and brown: target signals).}
			\label{fig:tsne}
		\end{figure}

		Fig.~\ref{fig:tsne} plots the t-distributed Stochastic Neighbour Embedding (t-SNE)~\cite{van2008visualizing} embedding of the source and target features in ``14-7 $\rightarrow$ 3-19'' task. We observe that features in source only are not clustered as clearly as ours, indicating that our method can adapt to signals in the target domain well.

		\section{Conclusion and Discussion}
		
		This paper proposed the cross-receiver RFFI problem and developed a  generalization error  bound for it. By gaining insight from the error bound, an adaptation model from RX-1 to RX-2 is developed by using domain alignment and adaptive pseudo-labeling techniques. Experimental results on real-world WiFi datasets demonstrate the effectiveness of the proposed method.

Before closing this paper, we would like to point out that the developed adaptation approach may  serve as  stepping-stone for solving more complex cross-receiver RFFI problems. For example, when there are multiple source receivers and/or multiple target receivers, we may  treat the multiple source/target receivers as a whole and directly apply the proposed adaptation approach to overall align the source and the target domains. However, this direct extension ignores distinct features within source/target domains. As a future work, apart from cross-receiver domain alignment, one  may  additionally take within-receiver  domain alignment into account to further improve the adaptation performance.

\appendices
\section{{Proof of Theorem~\ref{theorem}}} \label{appendix_proof_lemma}
 To prove Theorem~\ref{theorem}, the following lemma from~\cite[Theorem 2]{ben2010theory} is needed. 
\begin{lemma}[{\cite[Theorem 2]{ben2010theory}}]
	\label{target_error_bound}
	Let $\mathcal{H}$ be a hypothesis space of VC dimension $d$. If $\mathcal{S}$, $\mathcal{T}$ contains samples of size $N$ each, then for any $\rho \in (0,1)$, with probability at least $1-\rho$ (over the choice of the samples), for every $h \in \mathcal{H}$ the following inequality holds
	\begin{equation}
	\begin{aligned}
	{\epsilon}^t(h, f^t) \le ~&{\epsilon}^s(h, f^s) + \frac{1}{2}d_{\mathcal{H}\Delta\mathcal{H}}(\mathcal{S}, \mathcal{T}) + \Lambda.
	\end{aligned}
	\end{equation}
	where $\Lambda = 4 \sqrt{\frac{2d\log{2N}+\log{\frac{2}{\rho}}}{N}} + \Lambda^\star$ and $\Lambda^\star = \min_{h \in \mathcal{H}}{\epsilon^s(h, f^s) + \epsilon^t(h, f^t)}$ denotes the minimum combined risk on both domains.
\end{lemma}

With Lemma~\ref{target_error_bound}, we have
\begin{equation} \label{eq:lemma1_1}
\begin{aligned}
& | \epsilon_\mu^{s\&t}(h) - \epsilon^t(h, f^t) | \\
\le~& | \epsilon_\mu^{s\&t}(h) - \epsilon^t(h, \hat{f}^t) | + | \epsilon^t(h, \hat{f}^t) - \epsilon^t(h, f^t) | \\
=  ~& \mu | \epsilon^s(h, f^s) - \epsilon^t(h, \hat{f}^t) | + | \epsilon^t(h, \hat{f}^t) - \epsilon^t(h, f^t) | \\
\le~& \mu \left[ | \epsilon^s(h, f^s) - \epsilon^t(h, f^t) | + | \epsilon^t(h, f^t) - \epsilon^t(h, \hat{f}^t) | \right] \\
& + | \epsilon^t(h, \hat{f}^t) - \epsilon^t(h, f^t) | \\
=  ~& \mu | \epsilon^s(h, f^s) - \epsilon^t(h. f^t) | + (1+\mu) | \epsilon^t(h, \hat{f}^t) - \epsilon^t(h, f^t) | \\
\le~& \mu(\frac{1}{2} d_{\mathcal{H}\Delta\mathcal{H}}(\mathcal{S}, \mathcal{T}) + \Lambda) + (1 + \mu) |\epsilon^t(h, \hat{f}^t) - \epsilon^t(h, f^t)|.
\end{aligned}
\end{equation}

Then, we consider the link between the second term $|\epsilon^t(h, \hat{f}^t) - \epsilon^t(h, f^t)|$ and the accuracy of the pseudo-labeling function on the target domain $\epsilon^t(\hat{f}^t, f^t)$:

\begin{equation} \label{eq:lemma1_2}
\begin{aligned}
& | \epsilon^t(h, \hat{f}^t) - \epsilon^t(h, f^t) | \\
=~& \left| \mathbb{E}_{X^t \sim \mathcal{D}^t}\left[ \mathds{1}_{h(X^t) \neq \hat{f}^t(X^t)}  \right] - \mathbb{E}_{X^t \sim \mathcal{D}^t}\left[ \mathds{1}_{h(X^t) \neq f^t(X^t)}  \right] \right| \\
=~& \left| \mathbb{E}_{X^t \sim \mathcal{D}^t}  \left[ \mathds{1}_{h(X^t) \neq \hat{f}^t(X^t)} - \mathds{1}_{h(X^t) \neq f^t(X^t)}     \right]   \right|	\\
\leq~& \mathbb{E}_{X^t \sim \mathcal{D}^t} \left[  \left|    \mathds{1}_{h(X^t) \neq \hat{f}^t(X^t)} - \mathds{1}_{h(X^t) \neq f^t(X^t)}   \right|     \right] \\
=~& \mathbb{E}_{X^t \sim \mathcal{D}^t} \left[  \left|    \frac{1-h(X^t)\hat{f}^t(X^t)}{2} - \frac{1-h(X^t)f^t(X^t)}{2}   \right|     \right] \\
=~& \mathbb{E}_{X^t \sim \mathcal{D}^t} \left[  \left|    \frac{h(X^t)(f^t(X^t) - \hat{f}^t(X^t))}{2}   \right|     \right] \\
=~& \mathbb{E}_{X^t \sim \mathcal{D}^t} \left[ \mathds{1}_{\hat{f}^t(X^t) \neq f^t(X^t)}  \right] = \epsilon^t(\hat{f}^t, f^t).
\end{aligned}
\end{equation}

With \eqref{eq:lemma1_1} and \eqref{eq:lemma1_2}, we have
\begin{equation} \label{eq:middle_result}
\begin{aligned}
& | \epsilon_\mu^{s\&t}(h) - \epsilon^t(h, f^t) | \\
\le~& \mu(\frac{1}{2} d_{\mathcal{H}\Delta\mathcal{H}}(\mathcal{S}, \mathcal{T}) + \Lambda) + (1 + \mu) |\epsilon^t(h, \hat{f}^t) - \epsilon^t(h, f^t)| \\
\le~& \mu(\frac{1}{2} d_{\mathcal{H}\Delta\mathcal{H}}(\mathcal{S}, \mathcal{T}) + \Lambda) + (1 + \mu) \epsilon^t(\hat{f}^t, f^t).
\end{aligned}
\end{equation}
Eqn.~\eqref{eq:middle_result} reveals that for any 	hypothesis $h\in {\cal H}$, the weighted risk $\epsilon_\mu^{s\&t}(\hat{h})$ is close to the risk on the target domain $\epsilon^t(h, f^t)$ with high probability. With \eqref{eq:middle_result}, we are ready to characterize the performance gap between $\hat{h}$ and the optimal $h^\star$ on the target domain. Specifically, we  have the following result.
\begin{equation*} 
\begin{aligned}
&   \epsilon^t(\hat{h}, f^t) \\
\leq  &  \epsilon_\mu^{s\&t}(\hat{h}) + \frac{\mu}{2} d_{\mathcal{H}\Delta\mathcal{H}}(\mathcal{S}, \mathcal{T})  + (1 + \mu)  \epsilon^t(\hat{f}^t, f^t) + \mu \Lambda\\
\leq & \epsilon_\mu^{s\&t}(h^\star) + \frac{\mu}{2} d_{\mathcal{H}\Delta\mathcal{H}}(\mathcal{S}, \mathcal{T})  + (1 + \mu) \epsilon^t(\hat{f}^t, f^t) + \mu \Lambda \\
\leq & \epsilon^{t}(h^\star, f^t) +  \mu d_{\mathcal{H}\Delta\mathcal{H}}(\mathcal{S}, \mathcal{T}) + 2(1 + \mu)  \epsilon^t(\hat{f}^t, f^t)   + 2 \mu \Lambda, \\
\end{aligned}
\end{equation*}
where the first inequality is due to~\eqref{eq:middle_result} with $h = \hat{h}$, the second inequality is because that $\hat{h}$ is the minimizer of $\epsilon_\mu^{s\&t}(\hat{h})$, and the last inequality follows from~\eqref{eq:middle_result} with $h = h^\star$. \hfill $\blacksquare$

\section{Proof of Theorem~\ref{theorem:2}}
\label{appendix_proof_claim}

To proceed, let us introduce the following definitions. 
\begin{itemize}
	\item A function $f$ is $L$-Lipschitz if  $\| f({x}) - f({x'}) \| \le L \| {x} - {x'} \|$ holds  for $\forall {x}, {x'}\in {\rm Dom}f$. 
	\item A function $f$ if   $\nu$-smooth if  $\| \nabla f({x}) - \nabla f({x'}) \| \le \nu \| {x} - {x'} \|$  holds  for $\forall {x}, {x'}\in {\rm Dom}f$.
\end{itemize}
And we make the following assumptions.
\begin{assum}\label{assum}
	The estimate network $\sf T$, the feature extractor $\sf E$ and the classifier $\sf C$ are linear in their respective parameters $\theta_{\sf T}$, $\theta_{\sf E}$ and $\theta_{\sf C}$. 
\end{assum}
\begin{assum} \label{assum2}
	The network parameters  $\theta_{\sf T}$, $\theta_{\sf E}$ and $\theta_{\sf C}$ are respectively within some convex and compact sets.
\end{assum}
The  proof leverages on a general GAD convergence result developed in~\cite[Theorem 4.8]{lin2020gradient}, which is adapted below.
\begin{lemma} \label{lemma:stationary_point}
	Consider the minimax problem 
	\begin{equation}\label{eq:lemma_minimax}
	\min_{x \in \mathbb{R}^n}{\max_{ y \in {\cal Y}} {f({ x},{ y})}}
	\end{equation}
	where ${\cal Y}$ is a convex and bounded set with a diameter $D \ge 0$. Suppose $f(x,y)$ satisfies 1)  $f(x, \cdot)$ is concave for each $x$; 2) $f(\cdot, y)$ is $L$-Lipschitz for each $y \in {\cal Y}$, 3) $f(x,y)$ is $\nu$-smooth. Then, the iterates $\{(x^m,y^m)\}_{m=1,2,\ldots}$ generated by GAD 
	\begin{align*}
	x^{m+1} & = x^m  - \eta_x \nabla f(x^m, y^m) \\
	y^{m+1} & = y^m  + \eta_y \nabla f(x^{m}, y^m)
	\end{align*}
	with the step sizes $\eta_x = \Theta(\epsilon^4/ (\nu^3 L^2 D^2))$  and $\eta_y = \Theta(1/\nu)$ converge to  an $\epsilon$-stationary point of the minimax problem~\eqref{eq:lemma_minimax}.
\end{lemma}

Denote by $f(\theta_{\sf E}, \theta_{\sf C}, \theta_{\sf T})$ the objective function of the minimax problem~\eqref{eq:loss_all_eqv}. With Lemma \ref{lemma:stationary_point}, the proof boils down to verifying  that $f(\theta_{\sf E}, \theta_{\sf C}, \theta_{\sf T})$ satisfies the following three conditions under the Assumptions~\ref{assum} and \ref{assum2}:

\begin{enumerate}
	\item $f(\theta_{\sf E}, \theta_{\sf C}, \cdot)$ is concave for each $(\theta_{\sf E}, \theta_{\sf C})$,
	\item $f(\cdot, \cdot, \theta_{\sf T})$ is $L$-Lipschitz for each $\theta_{\sf T} \in {\cal Y}$,
	\item $f(\theta_{\sf E}, \theta_{\sf C}, \theta_{\sf T})$ is $\nu$-smooth.
\end{enumerate}
Since class weighting is independent of the parameters $\theta_{\sf E}, \theta_{\sf C}, \theta_{\sf T}$, it does not change the results of the analysis and will not be considered in subsequent proofs. Let us first rewrite $f$ as
\begin{equation} \label{eq:f_trans}
\begin{aligned}
f &= \tilde{\epsilon}_\mu^{s\&t}(\theta_{\sf E}, \theta_{\sf C} ) + \lambda\zeta(\theta_{\sf T},\theta_{\sf E}) \\
&= \frac{1}{n^s} \sum_{i=1}^{n^s} {\left( -\widetilde{y}_i^{s\top} \theta_{\sf C} \theta_{\sf E} x_i^s + \log{\sum_{j=1}^K}{\exp{(\theta_{\sf C} \theta_{\sf E} x_i^s)_j}}\right)} \\
&~~~~+ \frac{1}{n^t} \sum_{k=1}^{n^t} {\left( -\widetilde{y}_k^{t\top} \theta_{\sf C} \theta_{\sf E} x_k^t + \log{\sum_{g=1}^K}{\exp{(\theta_{\sf C} \theta_{\sf E} x_k^t)_g}}\right)} \\
&~~~~+ \frac{\lambda}{n^s} \sum_{m=1}^{n^s}{\theta_{\sf T} \theta_{\sf E} x_m^s} - \lambda \log { \left [\frac{1}{n^t} \sum_{q=1}^{n^t}{\exp{(\theta_{\sf T} \theta_{\sf E} x_q^t})} \right ]} \\
&= \frac{1}{n^s} \sum_{i=1}^{n^s} {\left( -\widetilde{y}_i^{s\top} \theta_{\sf C} \theta_{\sf E} x_i^s + {\rm lse}(\theta_{\sf C} \theta_{\sf E} x_i^s) \right)} \\
&~~~~+ \frac{1}{n^t} \sum_{j=1}^{n^t} {\left( -\widetilde{y}_j^{t\top} \theta_{\sf C} \theta_{\sf E} x_j^t + {\rm lse}(\theta_{\sf C} \theta_{\sf E} x_j^t) \right)} \\
&~~~~+ \frac{\lambda}{n^s} \sum_{m=1}^{n^s}{\theta_{\sf T} \theta_{\sf E} x_m^s} - \lambda \cdot{\rm lse}(\theta_{\sf T} \theta_{\sf E} X^t - \log{n^t}),
\end{aligned}
\end{equation}
where $X^t=[x_1^t, x_2^t, \cdots, x_{n^t}^t]$, $\widetilde{y}=onehot(y) \in \mathds{R}^{K}$ is a one-hot vector whose $y$-th element is $1$ and the rest of the elements are $0$, $\rm lse$ is the log-sum-exp function which is defined as
\begin{equation} \label{eq:def_lse}
{\rm lse}({\bm z}) = \upsilon^{-1} \log{\textstyle\sum_{i=1}^K {\upsilon z_i}}
\end{equation}
where ${\bm z} \in \mathds{R}^K$ and if not specified, $\upsilon=1$ in the remaining proof.

Since the log-sum-exp function is convex and monotonically increasing~\cite[Lemma 4]{gao2017properties}, it is easy to see that for fixed $(\theta_{\sf E}, \theta_{\sf C})$, $f$ is concave  in $\theta_{\sf T}$.

To prove the $L$-Lipschitz condition, we need the following facts:
\begin{fact}\label{fact:1}
	Let $g(\bm x, \bm y) = <\bm  x, \bm y>$ with $\bm x\in {\cal X}$ and $\bm y\in {\cal Y}$, where $<\cdot, \cdot>$ denotes the inner product, ${\cal X}$ and ${\cal Y}$ are compact  with diameters $D_x$ and $D_y$. Then, $g(\bm x, \bm y)$ is $L$-Lipschitz with $L = 2\max\{D_x, D_y\}$.
\end{fact}
\begin{fact}\label{fact:2}
	The log-sum-exp function ${\rm lse}(\bm x)$ is $L$-Lipschitz with $L = 1$.
\end{fact}
\begin{fact}\label{fact:3}
	The summation and composition of $L$-Lipschitz functions are still $L$-Lipschitz.
\end{fact}

The proof of Facts~\ref{fact:1}-\ref{fact:3} are given in Appendix~\ref{sec:appendix_fact_proof}. Under Assumption 2 and by Fact~\ref{fact:1}, we see that $-\widetilde{y}_i^{s\top} \theta_{\sf C} \theta_{\sf E} x_i^s$ is $L$-Lipschitz. By Facts~\ref{fact:2} and \ref{fact:3}, ${\rm lse}(\theta_{\sf C} \theta_{\sf E} x_i^s)$ is  $L$-Lipschitz. Similarly, we can show that the remaining terms in~\eqref{eq:f_trans} are $L$-Lipschitz. Therefore, by Fact~\ref{fact:3} their summation is $L$-Lipschitz.

To prove the $\nu$-smooth condition, we need the following facts:
\begin{fact}\label{fact:4}
	$g(\bm x, \bm y) = <\bm  x, \bm y>$ is $\nu$-smooth with $\nu  = 1$.
\end{fact}
\begin{fact}\label{fact:5}
	The log-sum-exp function ${\rm lse}(\bm x)$ is $\nu$-smooth with $\nu  = \upsilon$, where $\upsilon$ is defined in~\eqref{eq:def_lse}.
\end{fact}
\begin{fact}\label{fact:6}
	The summation and composition of  $\nu$-smooth functions are still  $\nu$-smooth.
\end{fact}

The proof of Facts~\ref{fact:4}-\ref{fact:6} are given in Appendix~\ref{sec:appendix_fact_proof}. Under Assumption~\ref{assum2} and by Fact~\ref{fact:4}, we see that $-\widetilde{y}_i^{s\top} \theta_{\sf C} \theta_{\sf E} x_i^s$ is $\nu$-smooth. By Facts~\ref{fact:5} and \ref{fact:6}, ${\rm lse}(\theta_{\sf C} \theta_{\sf E} x_i^s)$ is  $\nu$-smooth. Similarly, we can show that the remaining terms in~\eqref{eq:f_trans} are $\nu$-smooth. Therefore, by Fact~\ref{fact:6} their summation is $\nu$-smooth.

\section{Proof of Facts~\ref{fact:1}-\ref{fact:6}} \label{sec:appendix_fact_proof}
\label{appendix_proof_facts}
\subsection{Proof of Fact~\ref{fact:1}}
Given $\bm x, \bm x' \in {\cal X}$ and $\bm y, \bm y' \in {\cal Y}$, we have
\begin{align*}
& \| g(\bm x, \bm y) - g(\bm x', \bm y')\|  \\
= & \| <\bm x, \bm y> -<\bm x', \bm y> + <\bm x', \bm y> - <\bm x', \bm y'> \| \\
\leq & \| <\bm x, \bm y> -<\bm x', \bm y> \|  + \| <\bm x', \bm y> - <\bm x', \bm y'> \| \\
= & \| <\bm x- \bm x', \bm y>  \|  + \| <\bm x', \bm y- \bm y'> \| \\
\leq &\| \bm y \| \| \bm x- \bm x' \|  +  \| \bm x'\| \|  \bm y- \bm y'\| \\
\leq & D_y \| \bm x- \bm x' \| + D_x \|  \bm y- \bm y'\|\\
\leq & \max\{ D_x, D_y\} (\| \bm x- \bm x' \|  + \|  \bm y- \bm y'\|) \\
\leq & 2 \max\{ D_x, D_y\} \| (\bm x, \bm y) - (\bm x',\bm y') \|.
\end{align*}
\subsection{Proof of Fact~\ref{fact:2}}
Given $\bm x,\bm x'\in \mathbb{R}^K$, denote ${\rm U}(r)={\rm lse}(r\bm{x}+(1-r)\bm{x}'), r \in [0,1]$. According to the mean value theorem, $\exists \xi \in (0,1), \forall \bm{x},\bm{x}'$,
\begin{equation*}
{\rm U}'(\xi) = \frac{{\rm U}(1)-{\rm U}(0)}{1-0} = {\rm lse}(\bm{x}) - {\rm lse}(\bm{x}').
\end{equation*}
Then we have
\begin{equation*}
{\rm U}'(\xi) = \frac{\sum_{i=1}^K {e^{\xi x_i + (1-\xi) x'_i}} (x_i - x'_i)}{\sum_{j=1}^K {e^{\xi x_j + (1-\xi) x'_j}}}.
\end{equation*}
Therefore, 
\begin{equation*}
\begin{aligned}
\| {\rm lse}(\bm{x}) - {\rm lse}(\bm{x}') \| &= \left| \frac{\sum_{i=1}^K {e^{\xi x_i + (1-\xi) x'_i}} (x_i - x'_i)}{\sum_{j=1}^K {e^{\xi x_j + (1-\xi) x'_j}}} \right| \\
&\le \frac{\sum_{i=1}^K {e^{\xi x_i + (1-\xi) x'_i}} \left|x_i - x'_i\right|}{\sum_{j=1}^K {e^{\xi x_j + (1-\xi) x'_j}}} \\
&\le \frac{\sum_{i=1}^K {e^{\xi x_i + (1-\xi) x'_i}} \| \bm{x} - \bm{x}' \|_2}{\sum_{j=1}^K {e^{\xi x_j + (1-\xi) x'_j}}} \\
&= \| \bm{x} - \bm{x}' \|_2.
\end{aligned}
\end{equation*}

\subsection{Proof of Fact~\ref{fact:3}}
The proof of $L$-Lipschitz for summation is trivial. For the composition case, given $(\bm x, \bm y)$ and $(\bm x', \bm y')$, we have
\begin{equation*}
\begin{aligned}
& \| {\rm lse}(g(\bm x, \bm y)) - {\rm lse}(g(\bm x', \bm y')) \|\\
\leq &\| g(\bm x, \bm y) - g(\bm x', \bm y')\|  \\
\le & 2\max\{D_x, D_y\}  \| (\bm x, \bm y) - (\bm x',\bm y')  \|
\end{aligned}
\end{equation*}

\subsection{Proof of Facts~\ref{fact:4}-\ref{fact:6}}
Fact~\ref{fact:4} can be trivially proved by directly evaluating the difference of the gradient. Fact~\ref{fact:5} can be deduced by noting that the gradient of the log-sum-exp function is the softmax function. According to~\cite[Proposition 4]{gao2017properties}, we know that softmax function is $L$-Lipschitz with $L= \upsilon$, and consequently the log-sum-exp function is $\nu$-smooth with $\nu = \upsilon$. Fact~\ref{fact:6} is deduced as follows. Let $\bm z = (\bm x, \bm y)$ and $\bm z' = (\bm x', \bm y')$. We have
\begin{align*}
&  \| \nabla {\rm lse}(g(\bm x, \bm y))  - \nabla {\rm lse}(g(\bm x', \bm y')) \|  \\
= & \|  \nabla {\rm lse}(\bm z) \nabla g(\bm x, \bm y)  -   \nabla {\rm lse}(\bm z') \nabla g(\bm x', \bm y')  \| \\
= & \| \nabla {\rm lse}(\bm z) \nabla g(\bm x, \bm y) - \nabla {\rm lse}(\bm z') \nabla g(\bm x, \bm y) + \nabla {\rm lse}(\bm z') \nabla g(\bm x, \bm y)\\
&  -   \nabla {\rm lse}(\bm z') \nabla g(\bm x', \bm y')   \| \\
\leq &  \|  \nabla g(\bm x, \bm y) \|  \| \nabla {\rm lse}(\bm z) - \nabla {\rm lse}(\bm z')\| + \| \nabla {\rm lse}(\bm z') \| \times \\
&\| \nabla g(\bm x, \bm y)   -    \nabla g(\bm x', \bm y')   \| \\
\leq & \upsilon  \|  \nabla g(\bm x, \bm y) \| \| \bm z - \bm z'\| + \| \nabla {\rm lse}(\bm z') \| \| \bm z - \bm z'\| \\
= & \upsilon \| (\bm y, \bm x) \| \| \bm z - \bm z'\| + \| {\rm softmax}(\bm z') \| \| \bm z - \bm z'\| \\
\leq & \upsilon \sqrt{D_x^2 + D_y^2} \| \bm z - \bm z'\| + K \| \bm z - \bm z'\| \\
= & (\upsilon \sqrt{D_x^2 + D_y^2} +   K) \| \bm z - \bm z'\|
\end{align*}
where the third inequality is because the range space of  the softmax function is a probabilistic simplex with dimension $K$.
		
%\bibliographystyle{IEEEbib}
%\bibliography{refs}

\begin{thebibliography}{10}
\providecommand{\url}[1]{#1}
\csname url@samestyle\endcsname
\providecommand{\newblock}{\relax}
\providecommand{\bibinfo}[2]{#2}
\providecommand{\BIBentrySTDinterwordspacing}{\spaceskip=0pt\relax}
\providecommand{\BIBentryALTinterwordstretchfactor}{4}
\providecommand{\BIBentryALTinterwordspacing}{\spaceskip=\fontdimen2\font plus
\BIBentryALTinterwordstretchfactor\fontdimen3\font minus
  \fontdimen4\font\relax}
\providecommand{\BIBforeignlanguage}[2]{{%
\expandafter\ifx\csname l@#1\endcsname\relax
\typeout{** WARNING: IEEEtran.bst: No hyphenation pattern has been}%
\typeout{** loaded for the language `#1'. Using the pattern for}%
\typeout{** the default language instead.}%
\else
\language=\csname l@#1\endcsname
\fi
#2}}
\providecommand{\BIBdecl}{\relax}
\BIBdecl

\bibitem{zhang2023radio}
J.~Zhang, G.~Shen, W.~Saad, and K.~Chowdhury, ``Radio frequency fingerprint
  identification for device authentication in the {Internet of Things},''
  \emph{IEEE Communications Magazine}, 2023.

\bibitem{chen2019radio}
S.~Chen, H.~Wen, J.~Wu, A.~Xu, Y.~Jiang, H.~Song, and Y.~Chen, ``Radio
  frequency fingerprint-based intelligent mobile edge computing for internet of
  things authentication,'' \emph{Sensors}, vol.~19, no.~16, p. 3610, 2019.

\bibitem{tian2020identity}
Q.~Tian, Y.~Lin, X.~Guo, J.~Wang, O.~AlFarraj, and A.~Tolba, ``An identity
  authentication method of a {MIoT} device based on radio frequency ({RF})
  fingerprint technology,'' \emph{Sensors}, vol.~20, no.~4, p. 1213, 2020.

\bibitem{xu2021adaptive}
C.~Xu, F.~He, B.~Chen, Y.~Jiang, and H.~Song, ``Adaptive {RF} fingerprint
  decomposition in micro {UAV} detection based on machine learning,'' in
  \emph{ICASSP 2021-2021 IEEE International Conference on Acoustics, Speech and
  Signal Processing (ICASSP)}.\hskip 1em plus 0.5em minus 0.4em\relax IEEE,
  2021, pp. 7968--7972.

\bibitem{liu2022radio}
M.~Liu, C.~Liu, Y.~Chen, Z.~Yan, and N.~Zhao, ``Radio frequency fingerprint
  collaborative intelligent blind identification for green radios,'' \emph{IEEE
  Transactions on Green Communications and Networking}, 2022.

\bibitem{zhang2023data}
Z.~Zhang, L.~Yuan, F.~Zhou, and Q.~Wu, ``Data-and-knowledge dual-driven radio
  frequency fingerprint identification,'' \emph{IEEE Internet of Things
  Journal}, 2023.

\bibitem{merchant2019toward}
K.~Merchant and B.~Nousain, ``Toward receiver-agnostic {RF} fingerprint
  verification,'' in \emph{2019 IEEE Globecom Workshops (GC Wkshps)}.\hskip 1em
  plus 0.5em minus 0.4em\relax IEEE, 2019, pp. 1--6.

\bibitem{shen2022towards}
G.~Shen, J.~Zhang, A.~Marshall, R.~Woods, J.~Cavallaro, and L.~Chen, ``Towards
  receiver-agnostic and collaborative radio frequency fingerprint
  identification,'' \emph{arXiv preprint arXiv:2207.02999}, 2022.

\bibitem{devlin2018bert}
c.~J. Devlin, M.-W. Chang, K.~Lee, and K.~Toutanova, ``{Bert}: Pre-training of
  deep bidirectional transformers for language understanding,'' \emph{arXiv
  preprint arXiv:1810.04805}, 2018\color{black}.

\bibitem{ben2010theory}
S.~Ben-David, J.~Blitzer, K.~Crammer, A.~Kulesza, F.~Pereira, and J.~W.
  Vaughan, ``A theory of learning from different domains,'' \emph{Machine
  learning}, vol.~79, pp. 151--175, 2010.

\bibitem{lee2023weight}
E.~Lee, I.~Kim, and D.~Kim, ``Weight-based mask for domain adaptation,'' in
  \emph{ICASSP 2023-2023 IEEE International Conference on Acoustics, Speech and
  Signal Processing (ICASSP)}.\hskip 1em plus 0.5em minus 0.4em\relax IEEE,
  2023, pp. 1--5.

\bibitem{hanna2022wisig}
S.~Hanna, S.~Karunaratne, and D.~Cabric, ``Wisig: A large-scale {WiFi} signal
  dataset for receiver and channel agnostic {RF} fingerprinting,'' \emph{IEEE
  Access}, vol.~10, pp. 22\,808--22\,818, 2022.

\bibitem{vapnik1994measuring}
V.~Vapnik, E.~Levin, and Y.~Le~Cun, ``Measuring the {VC-dimension} of a
  learning machine,'' \emph{Neural computation}, vol.~6, no.~5, pp. 851--876,
  1994.

\bibitem{donsker1975asymptotic}
M.~D. Donsker and S.~S. Varadhan, ``Asymptotic evaluation of certain markov
  process expectations for large time, i,'' \emph{Communications on Pure and
  Applied Mathematics}, vol.~28, no.~1, pp. 1--47, 1975.

\bibitem{zhang2021flexmatch}
B.~Zhang, Y.~Wang, W.~Hou, H.~Wu, J.~Wang, M.~Okumura, and T.~Shinozaki,
  ``Flexmatch: {Boosting} semi-supervised learning with curriculum pseudo
  labeling,'' \emph{Advances in Neural Information Processing Systems},
  vol.~34, pp. 18\,408--18\,419, 2021.

\bibitem{lin2020gradient}
T.~Lin, C.~Jin, and M.~Jordan, ``On gradient descent ascent for
  nonconvex-concave minimax problems,'' in \emph{International Conference on
  Machine Learning}.\hskip 1em plus 0.5em minus 0.4em\relax PMLR, 2020, pp.
  6083--6093.

\bibitem{he2016deep}
K.~He, X.~Zhang, S.~Ren, and J.~Sun, ``Deep residual learning for image
  recognition,'' in \emph{Proceedings of the IEEE conference on computer vision
  and pattern recognition}, 2016, pp. 770--778.

\bibitem{ganin2016domain}
Y.~Ganin, E.~Ustinova, H.~Ajakan, P.~Germain, H.~Larochelle, F.~Laviolette,
  M.~Marchand, and V.~Lempitsky, ``Domain-adversarial training of neural
  networks,'' \emph{The journal of machine learning research}, vol.~17, no.~1,
  pp. 2096--2030, 2016.

\bibitem{saito2018maximum}
K.~Saito, K.~Watanabe, Y.~Ushiku, and T.~Harada, ``Maximum classifier
  discrepancy for unsupervised domain adaptation,'' in \emph{Proceedings of the
  IEEE conference on computer vision and pattern recognition}, 2018, pp.
  3723--3732.

\bibitem{liang2020we}
J.~Liang, D.~Hu, and J.~Feng, ``Do we really need to access the source data?
  source hypothesis transfer for unsupervised domain adaptation,'' in
  \emph{International conference on machine learning}.\hskip 1em plus 0.5em
  minus 0.4em\relax PMLR, 2020, pp. 6028--6039.

\bibitem{van2008visualizing}
L.~Van~der Maaten and G.~Hinton, ``Visualizing data using {t-SNE}.''
  \emph{Journal of machine learning research}, vol.~9, no.~11, 2008.

\bibitem{gao2017properties}
B.~Gao and L.~Pavel, ``On the properties of the softmax function with
  application in game theory and reinforcement learning,'' \emph{arXiv preprint
  arXiv:1704.00805}, 2017.

\end{thebibliography}

%\vspace{200pt}

\begin{IEEEbiography}[{\includegraphics[width=1in,height=1.25in,clip,keepaspectratio]{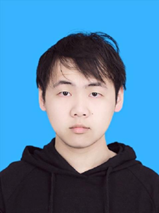}}]{Liu Yang}
	received the B.Eng. degree from the University of Electronic Science and Technology of China (UESTC), Chengdu, China, in 2020. He is currently pursuing the Ph.D. degree with the School of Information and Communication Engineering, UESTC. His current research interests include signal processing and radio frequency fingerprint identification.
\end{IEEEbiography}

\vspace{11pt}

\begin{IEEEbiography}[{\includegraphics[width=1in,height=1.25in,clip,keepaspectratio]{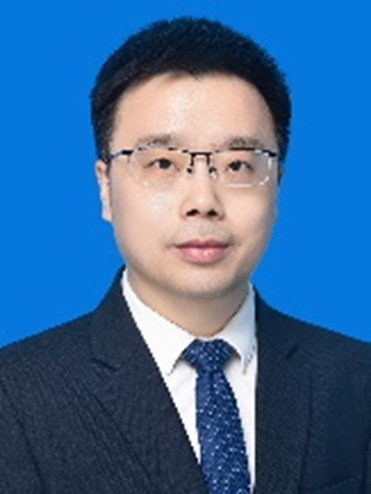}}]{Qiang Li}
 received the B.Eng. and M.Phil. degrees in communication and information engineering from the University of Electronic Science and Technology of China (UESTC), Chengdu, China, and the Ph.D. degree in electronic engineering from the Chinese University of Hong Kong (CUHK), Hong Kong, in 2005, 2008, and 2012, respectively. He was a Visiting Scholar with the University of Minnesota, and a Research Associate with the Department of Electronic Engineering and the Department of Systems Engineering and Engineering Management, CUHK. Since November 2013, he has been with the School of Information and Communication Engineering, UESTC, where he is currently a Professor. His recent research interests focus on machine learning and intelligent signal processing in wireless communications. He received the First Prize Paper Award in the IEEE Signal Processing Society Postgraduate Forum Hong Kong Chapter in 2010, a Best Paper Award of IEEE PIMRC in 2016, and the Best Paper Award of the IEEE SIGNAL PROCESSING LETTERS in 2016.
\end{IEEEbiography}

\vspace{11pt}

\begin{IEEEbiography}[{\includegraphics[width=1in,height=1.25in,clip,keepaspectratio]{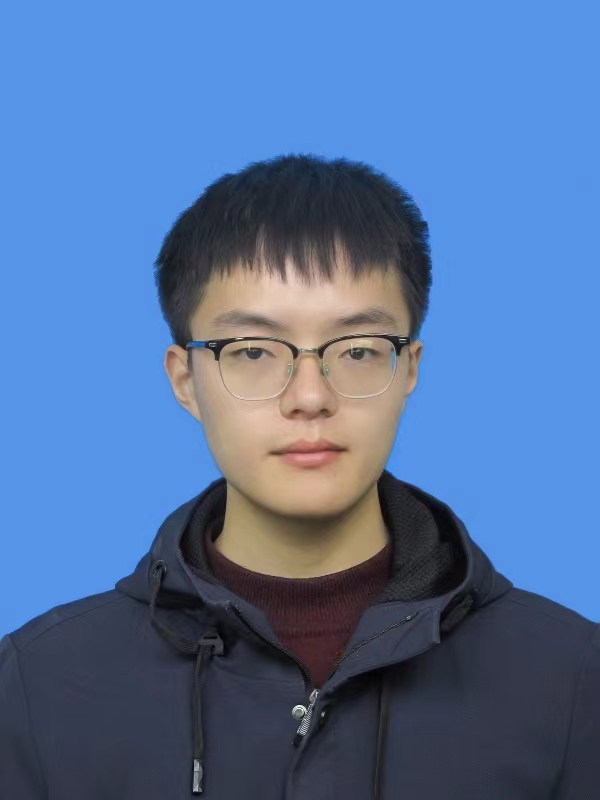}}]{Xiaoyang Ren}
	received the B.Eng. degree from Harbin Institute of Technology(HIT), Weihai, China, in 2021. He is currently pursuing the M.Eng. degree with the School of Information and Communication Engineering, University of Electronic Science and Technology of China (UESTC), Chengdu, China. His current research interests include the radio frequency fingerprint identification, domain adaptation and large language models.
\end{IEEEbiography}

\vspace{11pt}

\begin{IEEEbiography}[{\includegraphics[width=1in,height=1.25in,clip,keepaspectratio]{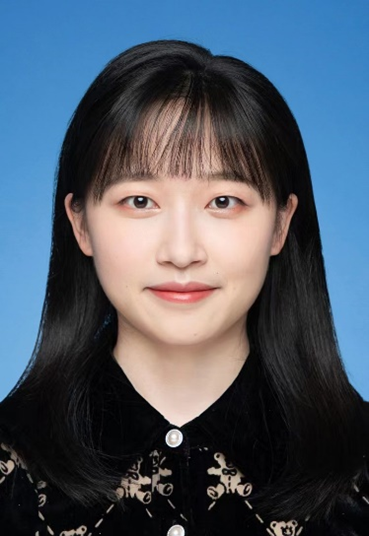}}]{Yi Fang}
	 received master’s degree in information and communication engineering from National University of Defense Technology. She is assistant research fellow at the Laboratory of Electromagnetic Space Cognition and Intelligent Control. Her research mainly focuses on signal processing. She has published 7 papers and obtained 4 patents.
\end{IEEEbiography}

\vspace{11pt}

\begin{IEEEbiographynophoto}{Shafei Wang}
	received the master's degree in signal and information processing from Beijing Institute of Technology (BIT), Beijing, China, in 1991. He is with the Laboratory of Electromagnetic Space Cognition and Intelligent Control, Beijing, China. His research interest includes signal processing.
\end{IEEEbiographynophoto}
		
	\end{document}